%% file: ricciardi.tex
\documentclass[useAMS,usenatbib]{mn2e}
\usepackage{graphicx,psfig,epsfig}

\def\lsim{\,\lower2truept\hbox{${<\atop\hbox{\raise4truept\hbox{$\sim$}}}$}\,}
\def\gsim{\,\lower2truept\hbox{${> \atop\hbox{\raise4truept\hbox{$\sim$}}}$}\,}

\begin{document}


\title[CCA on simulated polarized Planck data: errors estimation]{Correlated Component Analysis for diffuse component separation with error estimation on simulated Planck polarization data}

\author [Ricciardi, Bonaldi et al.] 
{S. Ricciardi$^{1}$, A. Bonaldi $^{2}$ , P. Natoli$^{3,4,5}$, G. Polenta$^{5,6}$, C. Baccigalupi$^{7}$, \newauthor 
E. Salerno$^{8}$, K. Kayabol$^{8}$, L. Bedini$^{8}$,  G. De Zotti$^{2,7}$
\vspace*{8pt} \\
$^{1}$ INAF/IASF, Sezione di Bologna, Via Gobetti, 101, I-40129 Bologna, Italy \\
$^{2}$ INAF, Osservatorio Astronomico di Padova, Vicolo dell'Osservatorio 5, I-35122 Padova,        Italy\\
$^{3}$ Dipartimento di Fisica, Universit\`a di Roma Tor Vergata, Via della Ricerca Scientifica 1, I-00133 Roma, Italy \\
$^{4}$ INFN, Sezione di Roma Tor Vergata, Via della Ricerca Scientifica 1, I-00133 Roma, Italy \\
$^{5}$ ASI Science Data Center, c/o ESRIN, I-00044 Frascati, Italy \\
$^{6}$INAF, Osservatorio Astronomico di Roma, via di Frascati 33, I-00040, Monte Porzio Catone, Italy \\
$^{7}$ SISSA, Via Beirut 2--4, I-34014 Trieste, Italy \\
$^{8}$ Istituto di Scienza e Tecnologie dell'Informazione, CNR, Area della ricerca di Pisa, via G. Moruzzi 1, I-56124 Pisa, Italy}

\maketitle
\begin{abstract}
We present a data analysis pipeline for CMB polarization experiments, running from multi-frequency maps to the power spectra. We focus mainly on component separation and, for the first time, we work out the covariance matrix accounting for errors associated to the separation itself. This allows us to propagate such errors and evaluate their contributions to the uncertainties on the final products.The pipeline is optimized for intermediate and small scales, but could be easily extended to lower multipoles. We exploit realistic simulations of the sky, tailored for the {\sc Planck} mission. The component separation is achieved by exploiting the Correlated Component Analysis in the harmonic domain, that we demonstrate to be superior to the real-space application \citep{bonaldi2006}. We present two techniques to estimate the uncertainties on the spectral parameters of the separated components. The component separation errors are then propagated by means of Monte Carlo simulations to obtain the corresponding contributions to uncertainties on the component maps and on the CMB power spectra. For the {\sc Planck} polarization case they are found to be subdominant compared to noise.
\end{abstract}
\section{Introduction}
\input{sec1.tex}
\section{Statement of the problem}\label{sec:problem}
\input{sec2.tex}
\section{Correlated Component Analysis (CCA)}\label{sec:method}
\input{sec3.tex}
\section{Simulated sky}\label{sec:dataset}
\input{sec4.tex}
\section{Estimating the mixing matrix parameters and errors with CCA}\label{sec:test}
\input{sec5.tex}

\section{Building full-sky maps of spatially-varying spectral indices}\label{sec:cca_maps}
\input{sec6.tex}

\section{Reconstruction of the components with GLS}\label{sec:gls}
\input{sec7.tex}
\section{CMB power spectrum estimation}\label{sec:pws}
\input{sec8.tex}
\section{Conclusions}
\input{sec9.tex}

\bibliographystyle{mn2e}
\bibliography{biblio}
\end{document}

%% file: sec1.tex
The Cosmic Microwave Background (CMB) radiation is a gold-mine of cosmological
information. While cosmology is entering its precision era, the target of CMB
experiments is shifting towards weak signals. The tiny amount of polarization
associated with the CMB anisotropy is undoubtedly one of the most intriguing -
and challenging - measurements of this kind. Several experimental efforts are
already pursuing the polarization of the CMB; many others will follow soon, as
the field is literally blossoming. The potential reward from this activity is
immense, since the CMB is thought to encode the solution to several long
standing puzzles in Cosmology and Fundamental Physics. While the largest
contribution to CMB polarization (so called E mode) arises due to the effect
of the scalar perturbations that also seed the large scale structure of the
Universe, theory predicts that a tiny part of the polarized signal is in the
form of (yet to be detected) B modes. On large angular scales, these modes
bear the imprint of the stochastic background of gravitational waves generated
during the inflationary phase of the Universe. At the same time, the amount of
power in B modes can be used to measure the energy scale of inflation, thus
probing particle physics. Furthermore, the different parity behavior of E and
B modes opens up the possibility to test for the breakdown of fundamental
symmetries. 

CMB anisotropies were first discovered by the COBE satellite
\citep{Smoot1992}, and higher resolution experiments detected the first
acoustic peaks \citep{DeBernardis1999,Hanany2000} in temperature. The E modes
were first detected from the ground by DASI \citep{Kovac2002}. 
Soon after, the WMAP satellite produced all sky data with a resolution down to
about 15 arcminutes \citep{Bennett2003}, impacting in particular on the
large scale E polarization of the CMB, and on the polarized foreground
properties \citep{Page2007}. The Planck satellite\footnote{Planck
{(http://www.esa.int/Planck)} is a project of the European Space Agency - ESA
- with instruments provided by two scientific Consortia funded by ESA member
states (in particular the lead countries: France and Italy) with contributions
from NASA (USA), and the telescope reflectors provided in a collaboration
between ESA and  scientific Consortium led and funded by Denmark.} 
 \citep{prelunch} is now measuring  the CMB anisotropy with an unprecedented accuracy.
Lately, experiments are focussing on the mapping of small scales total intensity anisotropies
\citep{Reichardt2009}, and of polarization \citep{2009ExA....23....5D}  
 with the ambitious goal of detecting the B modes for an interesting multipole range. The latter projects represent an extraordinary technological and scientific challenge, requiring a post-Planck, polarization dedicated satellite.

It must be clearly set forth that building high quality experiments is not the
only necessary condition for the CMB to disclose its secrets. It is also of
utmost importance to analyze the CMB data optimally in order to maximize the
information drawn from the data. First, one is looking for a tiny signal
buried under overwhelming instrumental noise, of both statistical and
systematic origin. Furthermore, the CMB is not the only signal on the sky:
other astrophysical sources exist, both compact and diffuse, that are powerful
emitters in the microwave band. Their emission can easily jeopardize
measurements of the CMB unless a very accurate separation of the astrophysical
components is achieved. 
Despite the large number of papers on the subject - from \cite{brandt1994} to
\cite{Stompor2009} and references therein, see \cite{DelabCardoso} for a
recent review - the handles used to achieve component separation are a few. On
one side, one can linearly combine the multifrequency maps in order to extract
from the data the most likely component scaling as a blackbody \cite[Internal
Linear Combination:][and references therein]{Bennett2003}, or can exploit the
statistical independence between components \cite[Independent Component
Analysis:][and references therein]{Stivoli2006,bet2009}, or adopt internal
template fitting procedures \citep{Martinez-Gonzalez2003}. On the other side,
one can exploit a partial knowledge of foregrounds, and in particular of their
frequency scalings, in order to parameterize them and measure their parameters
from multi-frequency maps, in the pixel domain
\citep{Stompor2009,Eriksen2006,bedini2005} or in the harmonic one
\citep{Vlad}. 
A crucially important still open issue is the estimation of errors on
separated maps and their propagation through all the steps of the analysis,
from the determination of spectral properties of the Galactic emissions to the
component maps and CMB power spectra. A correct  characterization of errors is
clearly essential for analyses of separated maps and for cosmological
parameter estimation. In this paper we present a pipeline for component
separation in polarization based on the Correlated Component Analysis
\citep[CCA,][]{bedini2005,bonaldi2006}, including error estimates. Two methods
for the estimation of errors on separated maps are presented and tested on
realistic simulations of Planck polarization data. 

The outline of the paper is the following. In $\S$\,\ref{sec:problem} we
formalize the component separation problem. In $\S$\,\ref{sec:method} we
describe our approach: the CCA technique both in the pixel and in the harmonic
domain, the associated error estimation and the reconstruction of the
individual components. In $\S$\,\ref{sec:dataset} we describe the simulations
used to test our pipeline. In $\S$\,\ref{sec:test} we assess the goodness of the
Galactic spectral indices estimated on sky patches, used in
$\S$\,\ref{sec:cca_maps} to build spatially varying spectral index maps. In
$\S$\,\ref{sec:gls} we present the component maps and the associated error
maps. Finally, in $\S$\,\ref{sec:pws} we present and discuss the estimated CMB
polarization power spectra. 

%% file: sec2.tex
The sky radiation, $\tilde{x}$, from the direction $\hat{r}$ at the frequency
$\nu$ results from the superposition of signals coming from $N_c$ different
physical processes $\tilde{s}_j$ : 

\begin{equation}
\tilde{x}(\hat{r},\nu)=\sum_{j=1}^{N_c}\tilde{s}_j(\hat{r},\nu).
\end{equation}
The signal $\tilde{x}$ is observed through a telescope, whose beam pattern can be modeled, at each frequency, as a shift-invariant point spread function $B(\hat{r},\nu)$.  For each
value of $\nu$, the telescope defocuses the physical radiation map by
convolving it with the kernel $B$.  The frequency-dependent convolved
signal is input to an $N_d$-channel measuring instrument, which
integrates the signal over frequency on each of its channels and adds
noise to its outputs.  The output of the measurement channel at a generic
frequency $\nu$ is 
\begin{equation}
x_\nu(\hat{r})=\int B(\hat{r}-\hat{r}',\nu')\sum_{j=1}^{N_c}t_\nu(\nu')
\tilde{s}_j(\hat{r}',\nu') d\hat{r}' d\nu'+n_\nu(\hat{r})\label{m0},
\end{equation}
where $t_\nu(\nu')$ is the frequency response of the channel and
$n_\nu(\hat{r})$ is the noise map.  The data model
in eq.~(\ref{m0}) can be simplified by virtue of the following
assumptions:
\begin{enumerate}
\item Each source signal is a separable function of direction and frequency:
\begin{equation}
\tilde{s}_j(\hat{r},\nu)=\bar{s}_j(\hat{r})f_j(\nu)\label{3}
\end{equation}
\item $B(\hat{r},\nu)$ is constant within the passband of the
measurement channel containing $\nu$.
\end{enumerate}
These two assumptions lead us to a new data model:
\begin{equation}
x_\nu(\hat{r})=B_\nu(\hat{r})*\sum_{j=1}^{N_c}h_{\nu j}\bar{s_j}(\hat{r}) +n_\nu(\hat{r})\label{m1},
\end{equation}
where $B_\nu(\hat{r})$ is the telescope beam pattern at the effective frequency
$\nu$, $*$ denotes convolution, and
\begin{equation}
h_{\nu j}=\int t_\nu(\nu')f_j(\nu')d\nu' \label{mixmat}.
\end{equation}
%
%
%
In vector form, we have
\begin{equation}
\mathbf{x}(\hat{r})=[\mathbf{B*H\bar{s}}](\hat{r})+\mathbf{n}(\hat{r}) \label{vect_m1}
\end{equation}
where $\mathbf{B}$ is a diagonal $N_{d}$-matrix whose elements are
$B_\nu(\hat{r})$, $\mathbf{H}$ is a constant $N_{d}\times N_{c}$
matrix whose elements are $h_{\nu j}$, $\mathbf{\bar{s}}$ is an
$N_{c}$-vector whose elements are $\bar{s_j}(\hat{r})$, and
$\mathbf{n}$ is an $N_{d}$-vector whose elements are $n_\nu(\hat{r})$.
The data model has thus become linear and convolutional, with known
point spread functions.

Equation (\ref{vect_m1}) can be translated to the harmonic domain
where, for each transformed mode, the linear convolutional model
becomes linear and instantaneous:
\begin{equation}
\mathbf{X}=\mathbf{H\tilde{B}}\mathbf{S}+\mathbf{N}\label{modhcca},
\end{equation}
where $\mathbf{X}$, $\mathbf{S}$, and $\mathbf{N}$ are the transforms
of $\mathbf{x}$, $\mathbf{\bar{s}}$, and $\mathbf{n}$, respectively,
and $\mathbf{\tilde{B}}$ is the transform of the matrix $\mathbf{B}$.

Let us now assume
that the beam patterns of the telescope are the same for
all the measurement channels, that is,
\begin{equation}
    B_\nu(\hat{r})=B(\hat{r}). \label{eqbeam}
\end{equation}
By virtue of this position, eq. (\ref{m1}) becomes
\begin{equation}
x_\nu(\hat{r})=\sum_{j=1}^{N_c}h_{\nu j} s_j(\hat{r})
+n_\nu(\hat{r})\label{inst_m1},
\end{equation}
with $s_j(\hat{r}) = [B*\bar{s}](\hat{r})$, or, in vector form,
\begin{equation}
\mathbf{x}=\mathbf{Hs}+\mathbf{n}.\label{modcca}
\end{equation}
Hereafter we drop the overbar from the symbol of the source vector.

It is worthwhile to note that eq.~(\ref{3}) comes from an
important assumption: the spectral properties of the astrophysical
sources are spatially uniform on the $n_p$ pixels
considered.  This assumption must be dealt with carefully because the Galactic
components are spatially varying, as we discuss better in
\S\,\ref{sec:dataset}.  Our strategy to overcome this difficulty is to apply
our method separately to  sky patches, where the foreground properties can be
safely assumed to be uniform. Generally, the assumption leading to
eq.~(\ref{eqbeam}), needed to build an instantaneous data model, is also not
realistic.
 A simple way to apply the model  [eq.~(\ref{modcca})] to a general case is to equalize the resolution of the instrumental
channels to the worse one.  For the harmonic domain this is not needed, so that the full instrumental resolution of each channel can be exploited.

\subsection{Strategy for component separation}
Equations~(\ref{modcca}) and (\ref{modhcca}) clearly show that the key
ingredient to estimate the source vector is the mixing matrix
$\mathbf{H}$. The Correlated Component Analysis (CCA),
described in the next section, gives an estimate of the
mixing matrix. This estimation could be
performed in the pixel domain (\S\, \ref{pixel_cca}) and in the
harmonic domain  (\S\,\ref{harmonic_cca}). Both methods exploit a
tessellation  of the data set and an estimation of the mixing matrix patch by
patch.
Once we have an estimate of $\mathbf{H}$, we can
compute a suitable matrix $\mathbf{W}$, sometimes called
reconstruction matrix, allowing us to obtain an estimate
$\mathbf{\hat{s}}$ of the components from the noisy data
$\mathbf{x}$:
\begin{equation}\label{recon}
\hat\mathbf{s}=\mathbf{Wx}.
\end{equation}
%
 with a reconstruction matrix $\mathbf{W}=\mathbf{W}(\mathbf{H})$.

 The reconstruction as in eq. (\ref{recon}) could be done in principle both in
pixel and in harmonic domain. As we will discuss in 
\S\, \ref{recon_components} we choose to perform the reconstruction in pixel
domain  as this technique allows us to account for spatially-varying mixing matrix in a more direct way.

%% file: sec3.tex
The CCA exploits a second-order statistics to estimate the mixing matrix from the statistics of data and noise. To reduce the number of unknowns, the mixing matrix is parameterized through a parameter vector $\mathbf{p}$, such that
$\mathbf{H}=\mathbf{H}(\mathbf{p})$.
To choose a suitable parameterization for $\mathbf{H}$ we use the fact
that its elements are proportional to the spectra of astrophysical
sources. As discussed in $\S$ \,\ref{sec:dataset}, this allows us to
reduce substantially the number of unknowns in the mixing matrix with
respect to the original $N_{d}\times N_{c}$ elements to be estimated.

\subsection{Pixel domain CCA}\label{pixel_cca}

From the pixel-space data model in
eq.~(\ref{modcca}), we easily derive the following second-order
statistics \citep{bedini2005}:
\begin{equation}
\mathbf{C_x}(\tau,\psi)=\mathbf{H}\mathbf{C_s}(\tau,\psi)
\mathbf{H}^T+\mathbf{C_n}(\tau,\psi). \label{cca_constr}
\end{equation}
The quantities $\mathbf{C_x}(\tau,\psi)$, $\mathbf{C_s}(\tau,\psi)$
and $\mathbf{C_n}(\tau,\psi)$ are the covariance matrices of data,
components and noise, respectively, computed for a generic
two-dimensional shift $(\tau,\psi)$.  For example, for the
data covariance matrix we have:
\begin{eqnarray}
\mathbf{C_x}(\tau,\psi)=\langle [\mathbf{x(\xi,\eta)}-\mu]
[\mathbf{x}(\xi+\tau,\eta+\psi)-\mu]^T \rangle \label{covmat},
\end{eqnarray}
where $\xi$ and $\eta$ are the coordinates of the two dimensional
space, $\tau$ and $\psi$ are increments in $\xi$ and $\eta$, $\langle
...\rangle$ denotes expectation under the appropriate joint
probability distribution, $\mu$ is the mean vector and the superscript
$T$ means transposition.  The application to sky patches is
straightforward, as we simply need to compute the covariances on a
suitable list of pixels.

In eq.~(\ref{cca_constr}), $\mathbf{C_x}(\tau,\psi)$
and $\mathbf{C_n}(\tau,\psi)$ can be computed from the data and the
known statistics of noise, while $\mathbf{H}$ and
$\mathbf{C_s}(\tau,\psi)$ are unknown.

Once we consider eq.~(\ref{cca_constr}) for a sufficient number of
shift pairs$(\tau,\psi)$, both $\mathbf{p}$ (and hence $\mathbf{H}$) and
$\mathbf{C_s}(\tau,\psi)$ can be estimated by minimizing the functional:

\begin{equation}
\mathbf{\Phi}[\mathbf{C_{s}},\mathbf{H}]\!=\! \sum_{\tau,\psi}\!\parallel \! \mathbf{H}\mathbf{C_{s}}(\tau,\psi)
\mathbf{H}^T- \mathbf{\hat{C}_{x}}(\tau,\psi)+ \mathbf{C_{n}} (\tau,\psi)\!
\parallel ^{2}\!. \label{cca_objective}
\end{equation}
\subsection{Harmonic-domain CCA}\label{harmonic_cca}
By dropping the equal beam assumption [eq~(\ref{eqbeam})] and relying on the data model
of eq.~(\ref{modhcca}) in the harmonic domain, we easily derive an equivalent
of eq.~(\ref{cca_constr}) in terms of power cross-spectra  \citep{proceedisti07}:

\begin{equation}
\widetilde{\mathbf{C}}_{\mathbf{x}}=\widetilde{\mathbf{B}}\mathbf{H}\widetilde{\mathbf{C}}_{\mathbf{s}}\mathbf{H}^T\widetilde{\mathbf{B}}^\dagger+\widetilde{\mathbf{C}}_{\mathbf{n}}
\label{hcca_constr}
\end{equation}
where $\widetilde{\mathbf{B}}$ is the transform of the matrix $\mathbf{B}$
introduced in the previous section, and the dagger superscript denotes
the adjoint matrix.  The matrices  $\widetilde{\mathbf{C}}_{\mathbf{x}}(\ell)$,
$\widetilde{\mathbf{C}}_{\mathbf{s}}(\ell)$ and
$\widetilde{\mathbf{C}}_{\mathbf{n}}(\ell)$, all depending on the multipole $\ell$, 
are the cross-spectra of the data, sources and noise, respectively. 
Formally, they are obtained by applying the spherical harmonic
transform to the covariance matrices in eq. (12). When working in
small patches, the cross-spectra are calculated by averaging
circularly the 2-dimensional discrete Fourier transform on the
rectangular grid.

If we reorder the matrices
$\widetilde{\mathbf{C}}_{\mathbf{x}}(\ell) -
\widetilde{\mathbf{C}}_{\mathbf{n}}(\ell)$ and
$\widetilde{\mathbf{C}}_{\mathbf{s}}(\ell)$ into vectors
$\mathbf{d}(\ell)$ and $\mathbf{c}(\ell)$, respectively, we can
rewrite eq. (15) as

\begin{equation}
\mathbf{d}(\ell) = \mathbf{H}_{\mathbf{k}}(\ell)\mathbf{c}(\ell),\label{fd_cca}
\end{equation}
where $\mathbf{H}_{\mathbf{k}}(\ell) =
[\widetilde{\mathbf{B}}(\ell)\mathbf{H}]\otimes[\widetilde{\mathbf{B}}(\ell)\mathbf{H}]$ and the symbol $\otimes$ denotes the Kronecker product.
$\mathbf{H}_{\mathbf{k}}(\ell)$ and $\mathbf{c}(\ell)$ contain the
unknowns of our problem (namely, the vector $\mathbf{p}$ and all
the components of the source cross-spectra), and
$\mathbf{d}(\ell)$ would be known if the data cross-spectra were
known.
In our case, we use binned cross-spectra,
$\widetilde{\mathbf{C}}_{\mathbf{x}}(\hat{\ell})$,
$\widetilde{\mathbf{C}}_{\mathbf{s}}(\hat{\ell})$ and
$\widetilde{\mathbf{C}}_{\mathbf{n}}(\hat{\ell})$, obtained by averaging the
transforms onto suitable spectral bins $D_{\hat{\ell}}$. Thus,
the cross-spectra of data can be estimated from the available data
samples:

\begin{equation}
\widetilde{\mathbf{C}}_{\mathbf{x}}(\hat{\ell}) = \frac{1}{M_{\hat{\ell}}}
\sum_{i,j
\in D_{\hat{\ell}}} \mathbf{X}(i,j)  \mathbf{X}^{\dag}(i,j) \label{dataspectrum}
 \end{equation}
where the pairs $(i,j)$ are the modes contained in the spectral
bin denoted by $D_{\hat{\ell}}$ in the transformed domain and $M_{\hat{\ell}}$
is the number of Fourier modes contained in the $\hat{\ell}$-th spectral
bin $D_{\hat{\ell}}$, $\hat{\ell}=1,\ldots,\hat{\ell}_{max}$. The set $D_{\hat{\ell}}$ can be any subset of the Fourier plane, such as an annular bin defined
by its mean radius and its thickness, which can easily
be related to a specific $\ell$-interval in the spherical
harmonic domain.
Since the left-hand side of eq. (16) can only be evaluated through
the empirical data cross-spectra of eq. (17), the data model is
affected by an estimation error $\epsilon(\hat{\ell})$, with covariance
matrix $\mathbf{N}_{\epsilon}(\hat{\ell})$:
\begin{equation}
\mathbf{d}(\hat{\ell}) = \mathbf{H}_{\mathbf{k}}(\hat{\ell})\mathbf{c}(\hat{\ell})+ \mathbf{\epsilon}(\hat{\ell}),\label{fd_cca_error}
\end{equation}
where $\mathbf{d}(\hat{\ell})$ is now computed using the observed data
cross-spectrum matrix.
Vectors $\mathbf{\epsilon}(\hat{\ell)}$ represent the differences between
the components of the actual data cross-spectrum matrix and those  evaluated
from the data through eq. (\ref{dataspectrum}), and, of
course, are ordered as vectors in the same way as $\mathbf{\tilde C_{x}}(\hat{\ell)}$.
This ordering induces the structure of their covariance matrices, $\mathbf{N_{\epsilon}}(\hat{\ell})$.
Let us express the mapping between the indices $j$ and $k$ in matrix $[\mathbf{\tilde C_{x}}(\hat{\ell}) - \mathbf{\tilde
C_{n}}(\hat{\ell)}]$ and the corresponding index $i$ in vector
$\mathbf{d}(\hat{\ell})$ as follows
\begin{equation}
d_{i}(\hat{\ell})=[\mathbf{\tilde C_{x}}(\hat{\ell}) - \mathbf{\tilde
C_{n}}(\hat{\ell})]_{j(i),k(i)}.
\label{mapping}
\end{equation}
With this position, if the estimation errors in $\mathbf{d}(\hat{\ell})$ are
uncorrelated to each other, the matrix $\mathbf{N_{\epsilon}}(\hat{\ell})$ is
diagonal, and its entries are given by

\begin{eqnarray}
N_{\epsilon, i i}(\hat{\ell})&=&2 \sigma_{j(i)}^{4}/M_{\hat{\ell}},  \,\,\,\,\,\,\,\,\, \mathrm{if
}\ j(i)=k(i)\nonumber\\
N_{\epsilon, i i}(\hat{\ell})&=&\sigma_{j(i)}^{2} \sigma_{k(i)}^{2}/M_{\hat{\ell}},
\, \mathrm{if}\  j(i) \neq k(i)
\label{error_covariance}
\end{eqnarray}
where $\sigma_{j}^{2}$ is the known variance of the $j$-th element in the
noise vector.
Details on
this derivation can be found in \citet{cnr.isti_2008-B4-007}.

Let us now exploit all the significant spectral bins, that is, let
us assume a suitable set $[1, \hat{\ell}_{max}]$ for $\hat{\ell}$, and rewrite eq.
(18) by stacking all the quantities for the available values of
$\hat{\ell}$:
\begin{equation}
\mathbf{d}_V=[\mathbf{d}(1), \mathbf{d}(2), ... \mathbf{d}(\hat{\ell}_{max})]^T,\label{data_vector}
\end{equation}
\begin{equation}
\mathbf{c}_V=[\mathbf{c}(1), \mathbf{c}(2), ... \mathbf{c}(\hat{\ell}_{max})]^T,\label{source_vector}
\end{equation}
\begin{equation}
\mathbf{\epsilon}_V=[\mathbf{\epsilon}(1), \mathbf{\epsilon}(2), ... \mathbf{\epsilon}(\hat{\ell}_{max})]^T,\label{error_vector}
\end{equation}
\begin{equation}
\mathbf{H_{kB}}=\left[\begin{array}{cccc}
    \mathbf{H_{k}}(1) & 0 & \ldots & 0  \\
    0 & \mathbf{H_{k}}(2) & \ldots & 0  \\
    0 & \ldots & \ldots & 0  \\
    0 & \ldots & \ldots & \mathbf{H_{k}}(\hat{\ell}_{max})
\end{array}\right].
\label{block_matrix}
\end{equation}
By these positions, and bearing in mind that  matrix $\mathbf{H_{kB}}$
is completely specified by the parameter vector $\mathbf{p}$, eq.
(\ref{fd_cca_error}) becomes
\begin{equation}
\mathbf{d}_V=\mathbf{H_{kB}}(\mathbf{p}) \cdot \mathbf{c}_V + \mathbf{\epsilon}_V,\label{fd_cca_error_V}
\end{equation}
which allows us to estimate the parameter vector and the source
cross-spectra by minimizing the functional:
\begin{eqnarray}\label{hcca_objective}
\!\!&&\mathbf{\Phi}[\mathbf{p},\mathbf{c}_V]=\\
\!\!&=&\!\!\!\![\mathbf{d}_V-\mathbf{H_{kB}(p)}\cdot \mathbf{c}_V]^T \mathbf{N_{\epsilon B}}^{-1} [\mathbf{d}_V-\mathbf{H_{kB}(p)}\cdot \mathbf{c}_V]\nonumber \\ &+&\lambda\mathbf{c}_V^T\mathbf{C}\mathbf{c}_V\nonumber
\end{eqnarray}
with
\begin{equation}\label{block_error_matrix}
\mathbf{N_{\epsilon B}}=\left[\begin{array}{cccc}
    \mathbf{N_{\epsilon}}(1) & 0 & \ldots & 0\\
    0 & \mathbf{N_{\epsilon}}(2) & \ldots & 0\\
    0 & \ldots & \ldots & 0\\
    0 & \ldots & \ldots & \mathbf{N_{\epsilon}}(\hat{\ell}_{max})
\end{array}\right]
\end{equation}
The term $\lambda\mathbf{c}_V^T\mathbf{C}\mathbf{c}_V$ is a
quadratic stabilizer for the source power cross-spectra whose
estimation is an ill-posed problem. The matrix
$\mathbf{C}$ must be suitably chosen and the parameter $\lambda$ must be
tuned to balance the effects of data fit and regularization in the
final solution.
The functional in
eq.~(\ref{hcca_objective}) can be considered as a negative joint
log-posterior for $\mathbf{p}$ and $\mathbf{c}_V$, where the first
quadratic form represents the log-likelihood, and the regularization
term can be viewed as a log-prior density for the source power
cross-spectra.

\subsection{Foregrounds spectral parameters error estimation}\label{sec:errors}
A standard, theoretically-based, method to estimate the errors in the recovery of the spectral parameters, $\Delta\mathbf{p}$, leading to an error in the mixing matrix $\Delta\mathbf{H}$, relies on the
analysis of the marginal probability for the parameters $\mathbf{p}$. In the next section we describe the formalization of this technique for harmonic domain CCA.
As an alternative to this method, we also implemented a simpler technique, exploiting the redundancy of solutions obtained for different sky patches.  This allows us to provide an error estimation also for pixel domain CCA, for which the formalization of the
marginal probability method is still under development, and to
cross-check the results obtained for the harmonic domain CCA.

\subsubsection{Marginal probability method for harmonic domain CCA}\label{sec:marginal}
To evaluate the estimation errors on $\mathbf{p}$, we can first derive,
from eq.~(\ref{hcca_objective}), the joint distribution of $\mathbf{p}$
and $\mathbf{c}_V$ and then marginalize it by integrating out
$\mathbf{c}_V$.

By developing the functional in eq.~(\ref{hcca_objective}) and taking
its exponential, it is easy to see that the marginal density of
$\mathbf{p}$ conditioned to $\mathbf{d}_V$ is given by
\begin{eqnarray}
p(\mathbf{p}|\mathbf{d}_V) \propto
\int_{}^{}e^{-\frac{1}{2}[\mathbf{c}^{T}_{V}(\mathbf{H_{kB}}^{T}\mathbf{N_{\epsilon
B}}^{-1}\mathbf{H_{kB}}+\lambda \mathbf{C})\mathbf{c}_{V}]} \times\\
e^{\mathbf{d}^{T}_{V}\mathbf{N_{\epsilon B}}^{-1}\mathbf{H_{kB}}\mathbf{c}_{V}}
\times e^{-\frac{1}{2}\mathbf{d}^{T}_{V}\mathbf{N_{\epsilon B}}^{-1}\mathbf{d}_{V}}
d\mathbf{c}_{V},\label{marginal_1}\nonumber
\end{eqnarray}
which, by dropping unessential constants, becomes \citep{cnr.isti_2008-B4-007}:
\begin{eqnarray}\label{marginal_2}
p(\mathbf{p}|\mathbf{d}_V) \propto
\sqrt{\mathrm{det}(\mathbf{H_{kB}}^{T} \mathbf{N_{\epsilon
B}}^{-1}\mathbf{H_{kB}}+ \lambda \mathbf{C})^{-1}} \times \\
e^{\frac{1}{2}(\mathbf{N_{\epsilon
B}}^{-1}\mathbf{H_{kB}})(\mathbf{H_{kB}}^{T} \mathbf{N_{\epsilon
B}}^{-1}\mathbf{H_{kB}}+ \lambda \mathbf{C})^{-1}(\mathbf{H_{kB}}^{T} \mathbf{N_{\epsilon
B}}^{-1}\mathbf{d_{V}})}.\nonumber
\end{eqnarray}
Studying the behavior of this marginal distribution is not difficult, since $\mathbf{p}$ is a
low-dimension vector (typically, it has two or three components).
From eq.~(\ref{marginal_2}), all the quantities related to the
parameter distributions can be evaluated and, possibly, exploited to
estimate all the relevant reconstruction errors.
As a quantitative index of the estimation errors $\Delta \mathbf{p}$
we simply assume the standard deviations evaluated from the
normalized-exponent marginal posterior.

\subsubsection{Spatial redundancy method}\label{sec:redundancy}
On a certain patch of the sky, we call $\mathbf{p}$ the true spectral
parameters and $\mathbf{\hat p}$ the estimated ones; the quantity we
want to estimate is $\Delta\mathbf{p}=\mathbf{p}-\mathbf{\hat p}$.
Let us suppose to have a set of  parameters estimated on a sample
of sky patches, widely overlapping to the considered one.
For this sample we call the true and the
estimated spectral indices $\{\mathbf{p}_j\}$ and $\{\mathbf{\hat
p}_j\}$ respectively.

For the moment we make a simplifying assumption, which we will relax
later: the spectral behavior of the components is uniform over the
area covered by the sample of patches we are considering.  In this
case we have:
\begin{equation}
\{\mathbf{p}_j\}=\mathbf{p},\label{err1}
\end{equation}
and the estimated parameters $\{\mathbf{\hat p}_j\}$ are
different measurements of $\mathbf{p}$.  Thus, we can estimate
$\Delta\mathbf{p}$ by comparing the actual estimation in the
considered patch $\mathbf{\hat p}$ with the expectation value of
$\mathbf{p}$ computed on the sample of patches:
\begin{equation}
\hat \Delta \mathbf{p}=\langle \{\mathbf{\hat p}_j\}\rangle-\mathbf{\hat p}.\label{err2}
\end{equation}
In a realistic situation, however, the spectral indices are spatially-varying on scales smaller than that of the patches. Thus, for each patch the true spectral indices have a certain distribution. 
Our assumption is that in this case eq.~(\ref{err1}), though not strictly true, still approximately holds. 
In fact, due to the autocovariance of the foreground signal, we do not expect discontinuities in the spectral index distribution in nearby regions. Moreover, as the patches are partially overlapping, their mean spectral indices are highly correlated.
Thus, we assume that eq.~(\ref{err2}) can still be used for an
approximate error estimation, provided that the patches are small and
overlapped enough. A test of this method using simulations is described in
\S\,\ref{sec:test}.

\subsection{Reconstruction of the components}\label{recon_components}

 Adopting the linear mixture model in eq.~(\ref{modcca}), we can find
 a solution of the component separation problem of the form of eq.
 (\ref{recon}).
By exploiting the CCA results on  sky patches we are able to
synthesize spatially varying spectral index maps  (see \S\,\ref{sec:cca_maps} for details).
Once we have an estimation of the mixing matrix $\mathbf{H}$, a
suitable choice for the reconstruction matrix $\mathbf{W}$ is the
Generalized Least Square solution (GLS):
\begin{equation}\label{gls}
\mathbf{W}=[\mathbf{H^{T}C_n^{-1}H}]^{-1}\mathbf{H^{T}C_n^{-1}} ,
\end{equation}
which only depends on the mixing matrix and on the noise covariance $\mathbf{C_n}$.
By applying eq.~(\ref{gls}) in pixel space we  are able to preserve  the spatial variability of the estimated mixing matrix.
This local information is very valuable, as
the spectral properties of the foregrounds depend on the line of
sight.
One disadvantage of using the pixel domain is that the
noise covariance $\mathbf{C_n}$ among different pixels does not vanish, so that the full calculation of eqs.~(\ref{gls}) and (\ref{recon}) is
very computationally demanding.  For full resolution maps, having
$\sim 10^7$ pixels, computing the full $\mathbf{C_n}$ is infeasible in practice, and we are forced to take into account only the diagonal noise covariance, i.e. to neglect
any correlation between noise in different pixels.  The full
calculation can be performed on low resolution maps having $\sim
10^3$ pixels. Such code, under development, will provide a full noise
covariance for the reconstructed low resolution CMB map and will naturally
couple CCA with low resolution power spectrum estimators such as Bol-Pol
\citep{bolpol}. 
Another disadvantage of the pixel domain is that to apply
the simple relation, eq.~(\ref{gls}), we need to equalize the beams, thus
losing part of the instrumental resolution.  This can be avoided
again at the cost of an increased computational complexity.
An iterative multi-resolution solver is under development, which will play perfectly with the harmonic CCA  fully exploiting  the native channel resolution.

\subsection{Errors on  component maps}\label{err_recon_components}
In general, each component map estimated through eq.~(\ref{recon}) is affected by both the instrumental noise and the residual contamination from the other components.  The former has covariance matrix
\begin{equation}
\rm{cov}(\hat\mathbf{s})=\mathbf{W}\mathbf{C_nW^{\dagger}}\label{covnoise}
\end{equation}
The latter will be estimated propagating the
errors on spectral parameters to the actual maps of
individual components. This is done by performing
separation using different realizations of the spectral
parameters from their associated distribution characterized
by the uncertainties described in this Section.
Following this idea, in  \S\,\ref{sec:gls} we  present an approximated
analysis and we  demonstrate that, in our simulations,
the error propagation from foregrounds spectral parameters to component maps is satisfying. Notice that
a straightforward, although computationally expensive,
extension of the present treatment could be exploited to
propagate covariances of the recovered errors on spectral
parameters, by simply computing them while conducting
spectral indices estimation in parallel for all patches.
This step might be crucial especially at large scales, where
foregrounds do have correlations which could impact
significantly on the separation errors in the form covariances
between the distribution of spectral parameters associated to
nearby patches.
This implementation is on the other hand beyond the scope of the
present paper that regards small and intermediate scales but we plan
to implement it for the pipeline of real data analysis.

%% file: sec4.tex
The main diffuse components present in the Planck channels are the CMB and the
emissions due to  our own Galaxy, namely thermal and anomalous dust,
synchrotron, and free-free. Free-free and anomalous dust emission are expected
to be essentially unpolarized. Since we deal here with 
polarization data, we are left with CMB, synchrotron and thermal dust emissions only.
We note, however, that the method described in this paper can be
straightforwardly applied also to total intensity data.  

\begin{figure}
\begin{center}
\includegraphics[width=5cm,angle=90]{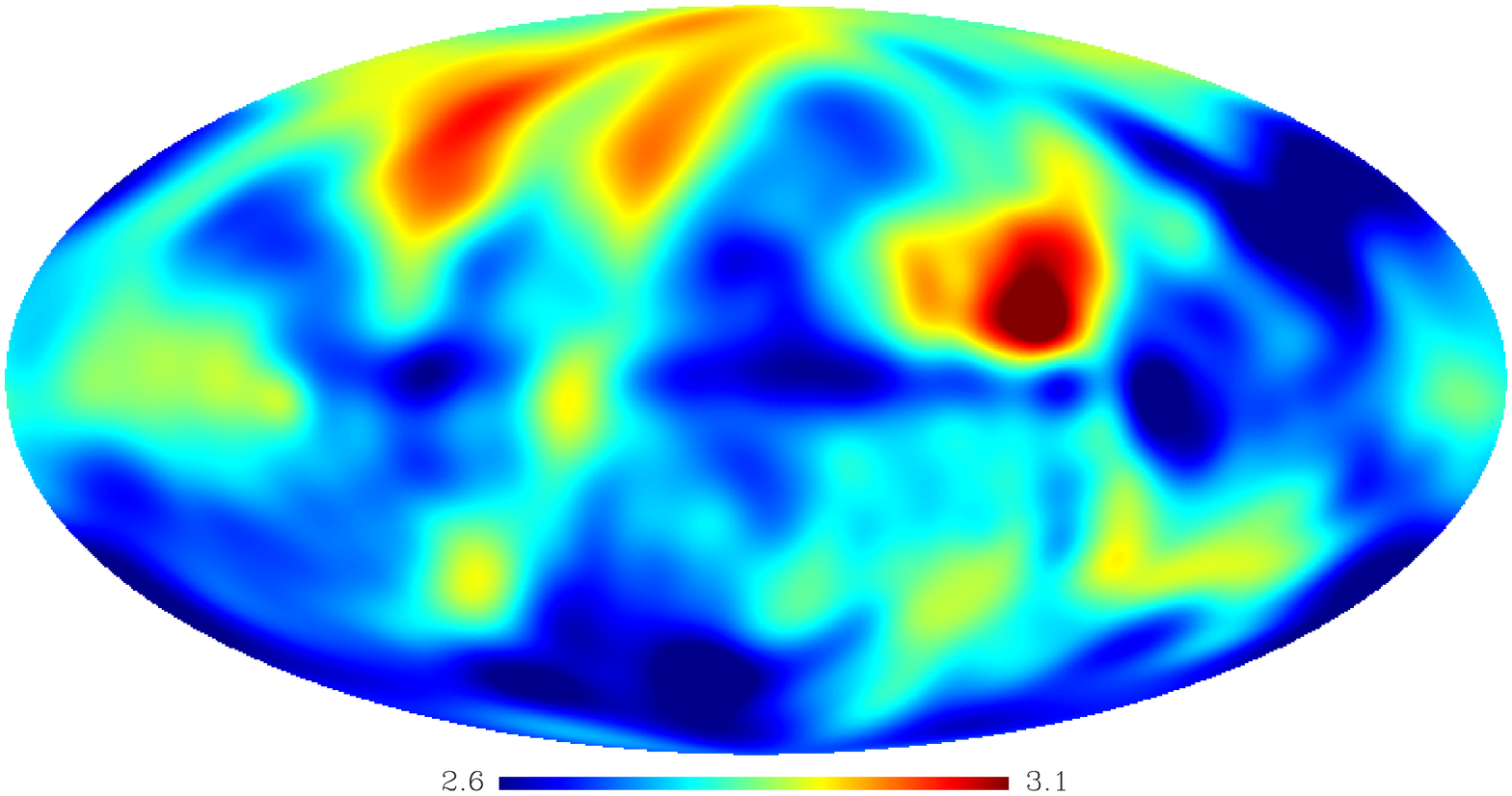}
\includegraphics[width=5cm,angle=90]{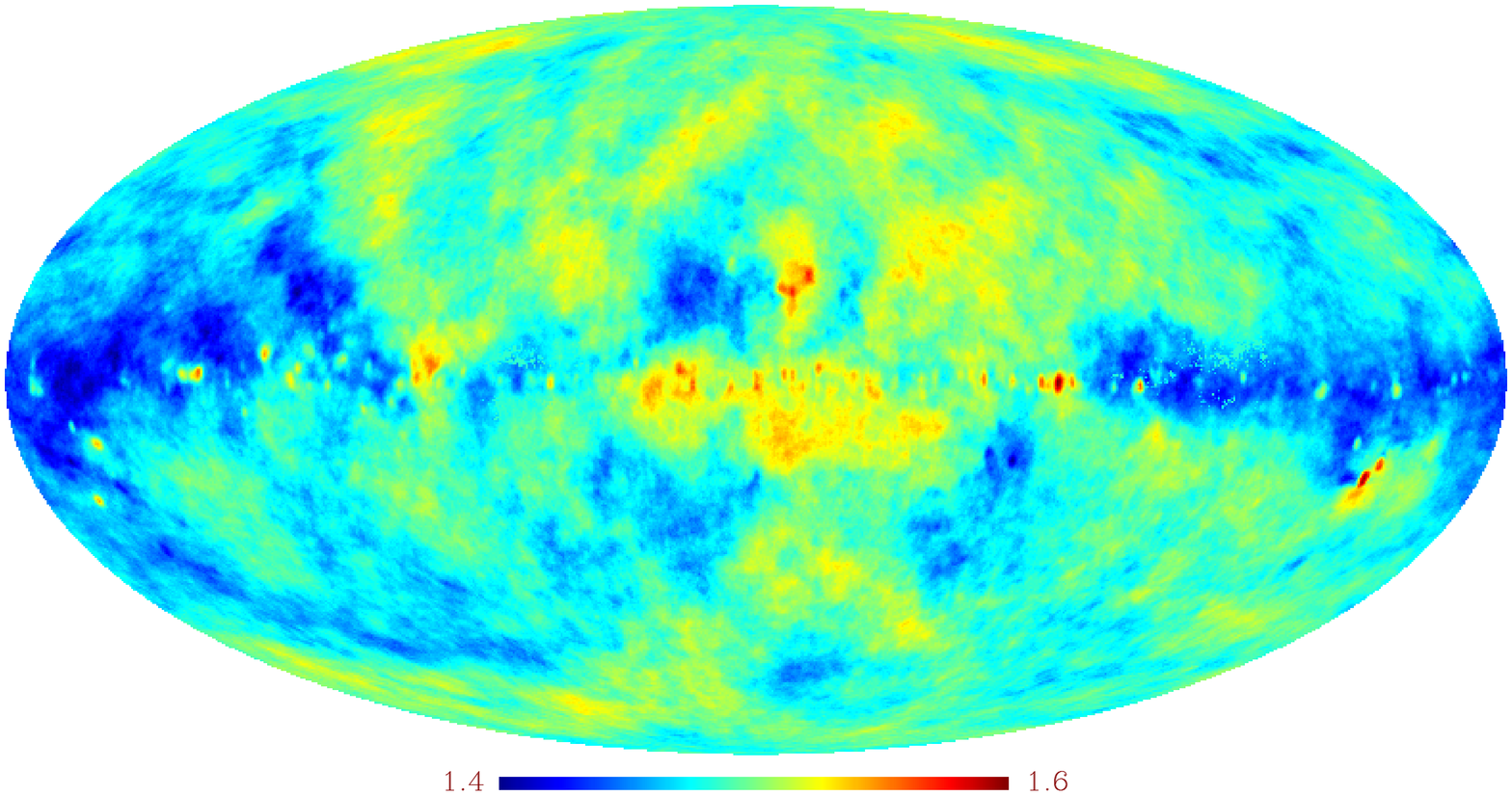}
\caption{Spectral index maps of synchrotron (top) and dust (bottom) components, used in the simulated sky}\label{fig:true_indices}
\end{center}
\end{figure}
Our simulation of the diffuse emission exploits all the available information
from existing public pre-Planck surveys. The simulated CMB map  is based on a
standard $\Lambda$CDM model consistent with WMAP 5-years cosmological
parameters, with a tensor to scalar ratio $r=0.1$. 
The rms fluctuations of the CMB are expressed in antenna temperature, $T_{\rm
  A,CMB}(\nu)$, whose frequency dependence writes:  
\begin{equation}\label{scalingCMB}
T_{\rm A,CMB}(\nu)\propto\frac{(h\nu/kT_{\rm CMB})^2\exp (h\nu/kT_{\rm CMB})}
{(\exp (h\nu/kT_{\rm CMB})-1)^2} \ ,
\end{equation}
where $h$ is the Planck constant, $k$ the Boltzmann constant,
$T_{\rm CMB}=2.726\,$K.

The simulation of the diffuse Galactic emissions is based on
\cite{miville2007} and implemented by the Planck working group on Component
Separation. The starting point towards building the synchrotron emission
template is an extrapolation of the 408 MHz map of \cite{haslam1982} from
which an estimate of the free-free emission has been removed. Due to the poor
resolution of the Haslam map (52 nominal arcmin) small structures have been
artificially added using the procedure  presented in  \cite{miville2007}. 
The fraction of polarization is derived from a model of the magnetic
field including  spiral and turbulent components based on WMAP 5yr results.
 On large scales ($> 5\,$deg) the polarisation angle is WMAP constrained; on smaller scales it relies on the Galactic magnetic field model. The spectrum, in antenna temperature, is assumed to follow a power law:
\begin{equation}\label{scalingSYNCHRO}
T_{\rm A,synch}(\nu)\propto \nu^{-\beta_s} \label{syn}\ ,
\end{equation}
where the synchrotron spectral index $\beta_s$ varies with the position on the sky. To this aim, we use the map of spectral indices given by \cite{giardino2002}. This map  shows structures on scales of up to 10 degrees, with $\beta_s$ varying from 2.5 to  3.2 (see the top panel of Fig.~\ref{fig:true_indices}).

The simulation of the polarized thermal dust emission is based on model 3 of
\cite{finkbeiner1999a}. This model extrapolates the $100\,\mu$m brightness map \citep{schlegel1998} assuming grey body spectra:
\begin{equation}\label{scalingDUST}
T_{\rm A,dust} (\nu)\propto \frac{\nu^{\beta_d+1}}{\exp
(h\nu/kT_{\rm dust})-1}, \label{dust}
\end{equation}
with $T_{\rm dust}=18\,$K and spatially varying emissivity index $\beta_d$, obtained from the  map of $100/240\,\mu$m flux density ratios published by \cite{schlegel1998}. The map of $\beta_d$ shows structure down to arcminute scales (see the bottom panel of Fig.~\ref{fig:true_indices}) with $\beta_d$ varying from 1.44 to 1.6. The polarized intensity is obtained multiplying the brightness map by a polarization fraction map extracted from WMAP 5yr data with the help of the model of the previously mentioned Galactic magnetic field model. In this simulation the polarization fraction is not frequency dependent.

Component maps have been produced at central frequencies of all {\sc Planck}
channels. They are then co-added and smoothed with nominal Gaussian beams (see
Table \ref{tab:planck}). 
A white inhomogeneous noise is synthesized using the block diagonal part of
the predicted noise covariance matrix given the {\sc Planck} nominal
integration time (14 months) and scanning strategy. The correlation between
noise in different Stokes parameters for each pixel has also been reproduced. 
This dataset has been complemented with the simulated WMAP 23 GHz map after 5  years of survey. This map has a resolution of 58 arcmin and the noise is simulated starting from the diagonal noise covariance provided by the WMAP team ({http://lambda.gsfc.nasa.gov}). This ancillary product is used to help tracing the low frequency foreground component for the CCA run, as described in \S\,\ref{sec:test}.
\begin{table}
\caption{Central frequencies, $\nu$, and angular resolutions for the {\sc Planck} channels considered in the present study } \label{tab:planck}
\begin{tabular}{|l|l|l|l|l|l|l|l|l|l|}
\hline
$\nu\,$(GHz)&30&44&70&100&143&217&353\\
\hline
FWHM (arcmin)&33&24&14&9.5&7.1&5.0&5.0\\
\hline
\end{tabular}
\end{table}

%% file: sec5.tex
We applied both pixel and harmonic domain CCA to the polarized {\sc Planck}
simulations  described in the previous section (7 frequencies from 30 GHz to
353 GHz), with and without the ancillary 23 GHz WMAP channel.
For the harmonic domain version we use the channel maps at their native
  resolution. In the application to polarized data we perform separated runs of the Q
and U maps so the $\mathbf{c}(\hat{\ell})$ in eq. (\ref{fd_cca_error}) simply
represent  cross-power spectra. We can evaluate the local mixing matrix
using both information  from Q and U maps. In practice, due to the lower
foregrounds signal in the U maps,
 the mixing matrix is basically defined by the Q maps analysis only.
The application of pixel-domain CCA requires that all the channel maps
have the same resolution. Thus we preliminarily degraded the maps to the
resolution of the lowest frequency channel (33 arcmin to work with {\sc
  Planck}  data only, 58 arcmin to include WMAP 23 GHz channel) convolving the
maps with a Gaussian beam.
The data model includes three diffuse components: CMB, synchrotron and thermal
dust, parameterized as in eqs.~(\ref{scalingCMB}), (\ref{scalingSYNCHRO}) and
(\ref{scalingDUST}) respectively. We estimated two free parameters, the
synchrotron and the dust indices, which were allowed to vary in the ranges
$2.3 \leq  \beta_s \leq 3.5$ and $ 1. \leq  \beta_d \leq 2.5$, while we kept
$T_{\rm dust}=18\,$K.
We note that we do not explicitly account for any modeling error in our
analysis, as the mixing matrix parametrization assumed by CCA exactly reflects
the one exploited for the data generation. Thus, the only systematic error is
given by the assumption that the spectral properties are constant within sky
patches.
Including the effect of an imperfect modeling is beyond the scope of this
paper, whose goal is to evaluate CCA performances \emph{per se}. The problem
was addressed in \cite{wmapbonaldi} where several models for the anomalous
emission were tested for the analysis of WMAP temperature data with CCA. In
any case, we believe that  such model uncertainties should be less severe in
polarization.

The choice of the patch size for the CCA run is a trade-off between the need
to have uniform foreground properties, which calls for small patches, and the
need to have enough statistics, which calls for bigger ones. The latter is
obviously related to the instrumental resolution of the data, as the
statistics is ultimately determined by the number of resolution elements in
the considered region of the sky. In the case of harmonic-domain CCA we have
the additional constraint of the maximum patch size allowed by the planar
approximation. However, the possibility of this version of CCA to handle
frequency-dependent instrumental beams, and thus to exploit the full
resolution, generally allows the use of smaller patches compared to pixel
domain CCA.
In this work we divide the sky in square patches for both pixel domain and
harmonic domain CCA. The pixel based version does not suffer of any constraint
regarding the shape of the patch. The current harmonic version instead can
work only on square regions. We adopt a patch size of $40^\circ \times
40^\circ$ for pixel-domain CCA, of $30^\circ \times 30^\circ$ for
harmonic-domain CCA. The centers of the patches are equally spaced in latitude
and longitude with shifts of $3^\circ$, up to a maximum central latitude of
$\pm 30^\circ$ so that we have 2520 patches in the sky.
This helps us to build a smooth spectral index map and allows a more localized
spectral index estimation, as described in \S\,\ref{sec:cca_maps}. Moreover
this provides the redundancy needed for error estimation as described in
\S\,\ref{sec:redundancy}, for which we considered samples of patches
overlapping by more than 60\%. We note however that this purely geometrical
partition of the sky is not driven by any astrophysical reason. If we could
achieve a partition that maximizes the uniformity of the spectral properties
within each patch, the CCA estimation would be more accurate. The full run
took $\sim$ 10 hours on a parallel machine at NERSC using 120
processors. Convergence has been reached in  $\sim 80\%$ of the patches for
pixel domain CCA and $\sim 90\%$ of the patches for harmonic domain CCA.

\subsection{Evaluation of the results}

\subsubsection{Actual errors on parameter estimation}

Ideally, to evaluate the quality of CCA results we should compare the true
synchrotron and dust spectral indices with the estimated ones for each
position in the sky. However, CCA only provides spectral indices per patch,
and the true spectral indices in general vary within the patch. Thus we
computed:
\begin{equation}
\Delta \mathbf{p}=\mathbf{\bar p}-\mathbf{\hat p},\label{true_error}
\end{equation}
where $\mathbf{\bar p}=[\bar \beta_s,\bar \beta_d]$ are the flux weighted
averages of the true parameters over the patch  and $\mathbf{\hat p}=[\hat
\beta_s,\hat \beta_d]$ are the estimates provided by CCA.
This choice for the ``true'' spectral indices per patch seems to be the most
appropriate because the CCA results mostly reflect the spectral properties of
the brightest foreground structures in the considered patch. Such bright
structures are obviously those that need to be more accurately removed to
produce a clean CMB map. In any case, a systematic difference is expected
because the ``true'' mean spectral indices and the ones recovered by CCA have
somewhat different meanings.

\begin{figure*}
\begin{center}
\includegraphics[width=10cm,angle=90]{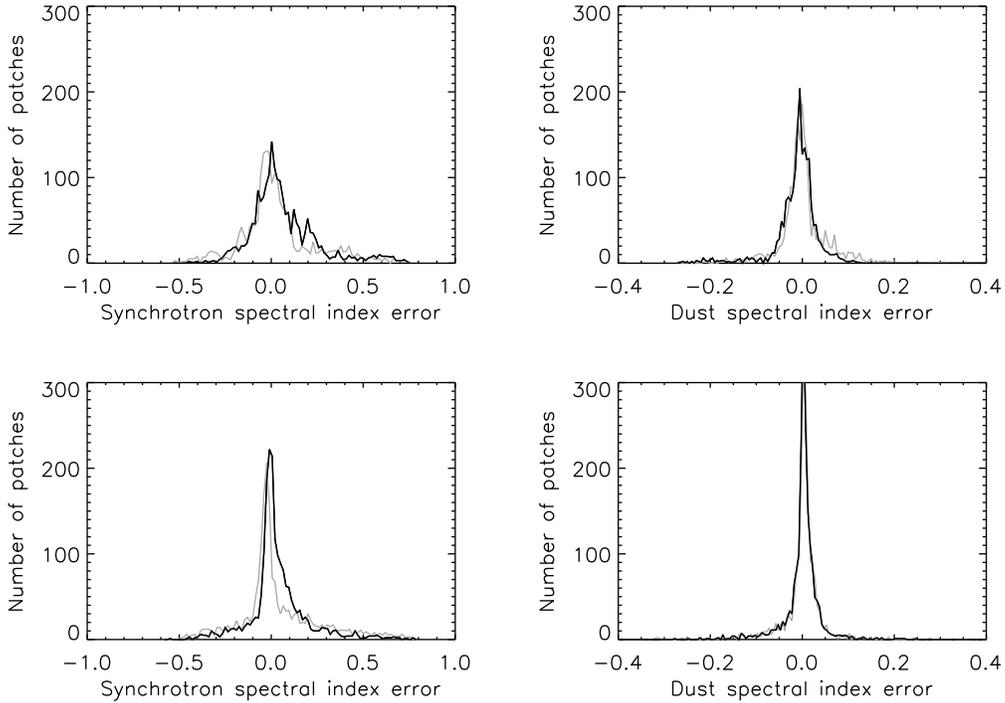}
\caption{Distribution of errors in the recovery of synchrotron (left) and dust (right) spectral indices for the full sample of sky patches using only the simulated {\sc Planck} maps (grey) and adding the simulated WMAP 23 GHz map (black). Top: pixel domain CCA; bottom: harmonic domain CCA}\label{fig:istogrammi_full}
\end{center}
\end{figure*}

In Fig.~\ref{fig:istogrammi_full} we show for each parameter and for both
pixel domain and harmonic domain CCA the distribution of ``true'' errors for
the full sample of patches obtained from simulated {\sc Planck} Q maps with
and without the simulated WMAP 23 GHz channel.
Table~\ref{tab:distr} reports the offset (from zero) of the mean and the
standard deviation of a gaussian fit computed for each histogram. We note that
the offset is in some cases comparable with the rms; this systematic error is
induced by the spatial variability of the true indices, as  mentioned
above. We verified that, once we adopt spatially-uniform spectral indices for
the data generation ($\beta_s = 2.9$ and  $\beta_d = 1.7$), the offset
disappears while the width of the distributions is almost unchanged.
Errors in the recovery of the synchrotron spectral index are bigger than those for the dust.
This is due to the fact that the dust component is much better traced by the
frequency coverage of {\sc Planck}. The inclusion of the WMAP 23 GHz channel
almost cancels the offset from zero of the mean error on the synchrotron
spectral index for harmonic domain CCA, while slightly increasing the rms
value (see Table~\ref{tab:distr}). For pixel domain CCA the advantage of a
broader frequency range yielded by the inclusion of the 23 GHz map is more
than compensated by the degradation of the resolution of the whole data set;
as a result, both the mean offset and the rms value of errors somewhat
increase.
In the following we only consider the results obtained with {\sc Planck} maps
alone for pixel domain CCA, and with {\sc Planck}\,$+$\,23\,GHz data for
harmonic domain CCA. The rms errors in the synchrotron and dust spectral
indices are $0.08$ and $0.019$ respectively for pixel domain CCA, $0.045$ and
$0.009$ respectively for harmonic domain CCA. Thus the harmonic domain CCA
performs better than the pixel domain version.

\begin{table}
\begin{center}
\caption{Offsets of the mean and rms values for the error distributions in Fig.~\protect\ref{fig:istogrammi_full} (Planck dataset/Planck + 23 GHz data set)}
 \label{tab:distr}
\begin{tabular}{|l|l|c|c|}
\hline
&&offset&rms\\
\hline
Synchr.&Pixel CCA&-0.009/0.016&0.080/0.107\\
&Harmonic CCA&-0.027/0.002&0.036/0.045\\
\hline
\hline
Dust&Pixel CCA&-0.005/-0.004&0.019/0.022\\
&Harmonic CCA&0.005/0.003&0.010/0.009\\
\hline
\end{tabular}
\end{center}
\end{table}

\begin{figure*}
\begin{center}
\includegraphics[width=7cm,angle=0]{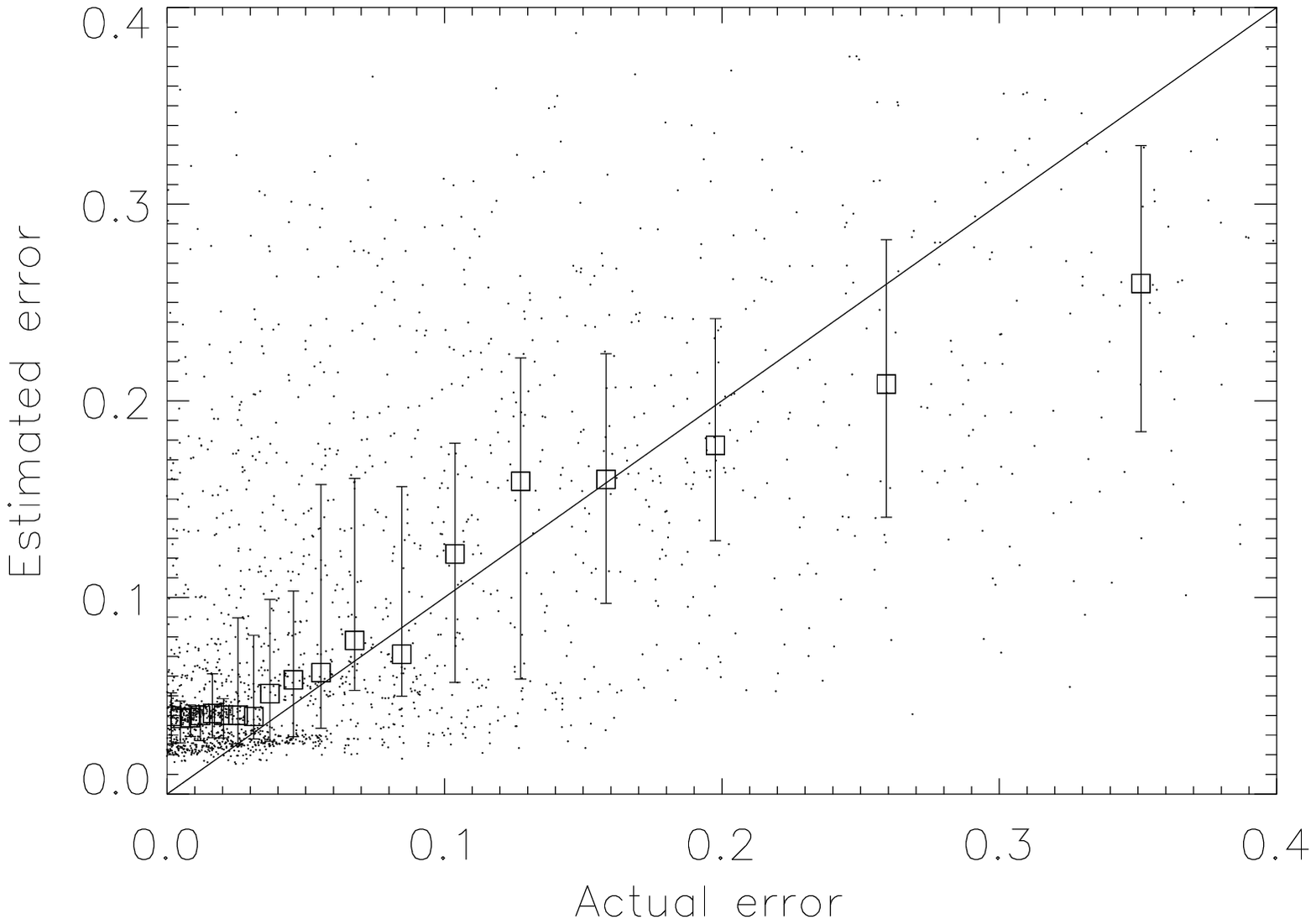}
\includegraphics[width=7cm,angle=0]{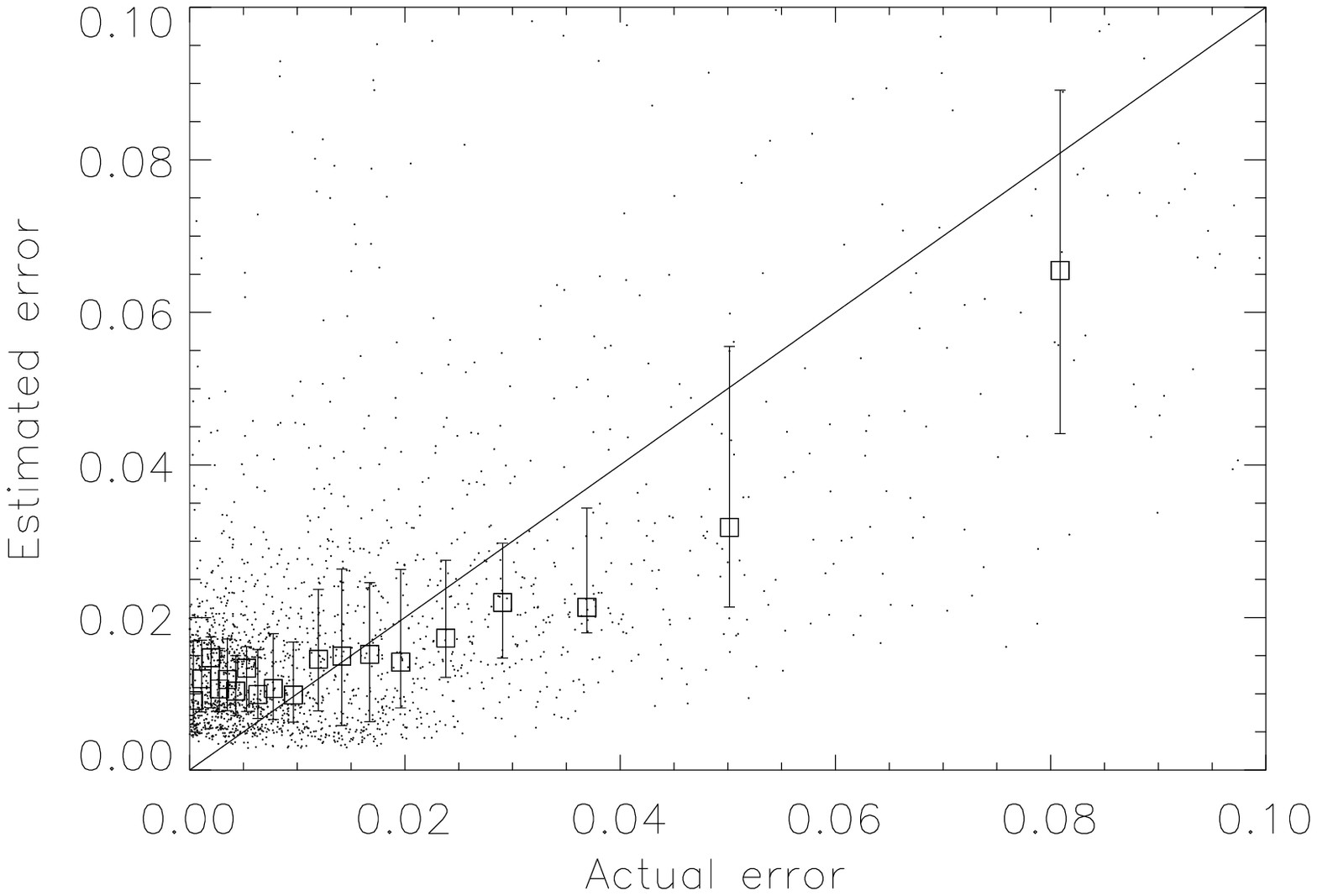}
\caption{Absolute values of estimated vs actual error on synchrotron (left)
  and dust spectral index (right) for the harmonic domain CCA; error
  estimation through the marginal distribution method. Dots are the values for
  individual patches, squares and error bars are the median values and
  quartiles computed for groups of 100 points, respectively}. 
\label{fig:stim_vs_true_marginals}
\end{center}
\end{figure*}

\subsubsection{Actual versus estimated errors}\label{sec:5.1.2}

As illustrated by Fig.~\ref{fig:stim_vs_true_marginals} 
the errors estimated via the marginal distribution method
(\S\,\ref{sec:marginal}.) are well correlated to the ``true'' errors above
$\Delta \beta_s\simeq 0.05$ or $\Delta \beta_d\simeq 0.01$. When the estimated
errors are small, the ``true'' errors are small as well, although the two
quantities are not correlated. This is due to the effect of noise which
hampers the calculation of the marginal probability distribution with
sufficient accuracy. The largest errors are moderately underestimated; this is
not particularly worrisome however, since large errors correspond to regions
where the foreground signals are very weak and can be removed with sufficient
accuracy adopting the mean parameter values determined in the rest of the
sky. 
The spatial redundancy error estimation method (\S\,\ref{sec:redundancy})
applied to both pixel and harmonic CCA yields estimated errors systematically
lower than the ``true'' ones. However, the latter are well matched by the
third quartiles of the distribution of estimated errors. As expected, the
spatial redundancy method performs better for the harmonic than for the pixel
domain CCA, thanks to the higher spatial resolution ($30^\circ \times
30^\circ$ versus $40^\circ \times 40^\circ$ patches). 

\begin{figure}
\begin{center}
\includegraphics[width=6.5cm,angle=90]{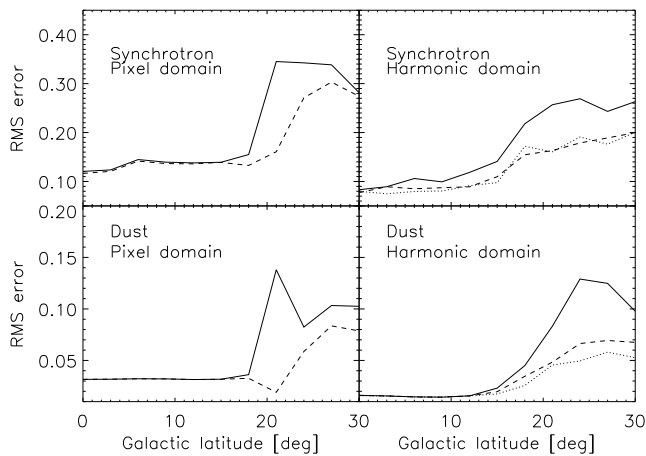}
\caption{Rms errors in the estimation of spectral indices as a function of Galactic latitude. Solid lines: full sample; dashed lines: final sample obtained with spatial redundancy error estimation method; dotted lines: final sample obtained with marginal distribution error estimation method. The errors increase at high Galactic latitudes, where the foreground signal are weak and there removal is therefore not a critical issue.}
\label{fig:err_vs_lat}
\end{center}
\end{figure}

Despite the non-idealities mentioned above, errors estimates produced by both
methods are good enough to be exploited in the next step of our pipeline.  In
the following analysis we will restrict ourselves to values of each spectral
index $\beta$ within the range $|\hat \Delta \beta|<\mu_{\hat
  \Delta\beta}+\sigma_{\hat \Delta\beta}$, where $\hat \Delta\beta$ is the
estimated error on $\beta$ and $\mu_{\hat \Delta\beta}$ and $\sigma_{\hat
  \Delta\beta}$ are the mean and the standard deviation of the estimated
errors over all the patches. This  set of values constitutes the \emph{final
  sample}. Figure~\ref{fig:err_vs_lat} compares the rms values of the actual
estimation errors as a function of Galactic latitude for both the full and the
final sample, showing that indeed this procedure removes the most discrepant
values. As already pointed out by \cite{wmapbonaldi}, the performance of CCA
is highly dependent on the intensity of the foreground signal in the
considered patch. Estimations of the spectral parameters are better close to
the Galactic plane and worsen with increasing Galactic latitude where,
however, the foreground signals are anyway very weak and their removal is
therefore not a critical issue. Our method allows us to flag the worse
estimates and to replace them with average values determined in the rest of
the sky. 

%% file: sec6.tex
\begin{figure}
\begin{center}
\includegraphics[width=8cm]{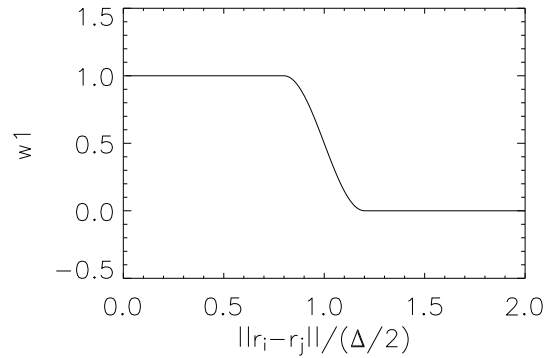}
\caption{Dependence of the weight w1 on the distance from the patch center normalized to half of the patch size $\Delta$}
\label{fig:w1}
\end{center}
\end{figure}

The  information on spectral parameters provided by CCA for sky patches can be
combined to build full-sky maps of the estimated, spatially-varying spectral
indices, and the associated estimation error maps. We use the Healpix
\citep{healpix} pixelization, with resolution parameter NSIDE=64,
corresponding to pixel areas of about 1 square degree. 
Let us consider the component $\rm p$ (e.g. the synchrotron spectral index) of
the parameter vector $\mathbf{p}$. Due to the partial overlap of sky patches,
the $j$-th pixel of the Healpix map, whose position on the sky is centered at
$\mathbf{r_j}$, belongs to several different patches for each of which we have
obtained an estimate of $\rm p$. We can therefore compute a weighted mean of
the values of the parameter  $\rm p_\mathbf{j}$ and of the error $\Delta \rm
p_\mathbf{j}$ as follows: 
\begin{eqnarray}
\rm p_{j}&=&\frac{\sum_{i=1}^N {\rm p}(\mathbf{r}_i)\cdot w1(\mathbf{r}_i,\mathbf{r}_j)\cdot w2(\mathbf{r}_i)}{\sum_{i=1}^N w1(\mathbf{r}_i,\mathbf{r}_j)\cdot w2(\mathbf{r}_i)},\\\label{map1}
\Delta \rm p_{j}&=&\frac{\sum_{i=1}^N\Delta {\rm p}(\mathbf{r}_i)\cdot w1(\mathbf{r}_i,\mathbf{r}_j)}{\sum_{i=1}^N w1(\mathbf{r}_i,\mathbf{r}_j)},\label{map2}
\end{eqnarray}
where the sum is over all patches in the final sample, defined in \S\,\ref{sec:5.1.2}, and $w1$ and $w2$ are weight functions. The former, $w1(\mathbf{r}_i,\mathbf{r}_j)$, depends on the distance between the $j$-th pixel and the center of the $i$-th patch,  $||\mathbf{r}_i-\mathbf{r}_j||$, normalized to the half size of the patch, $\Delta/2$. We used the function shown in Fig.~\ref{fig:w1}, equal to 1 for $||\mathbf{r}_i-\mathbf{r}_j|| \ll \Delta/2$, and equal to 0 for $||\mathbf{r}_i-\mathbf{r}_j|| \gsim \Delta/2$.
The weight $w2(\mathbf{r}_i)=w2(\Delta \rm p(\mathbf{r}_i))$ depends on the estimated error for the $i$-th patch trough the relation:
\begin{equation}
w2(\mathbf{r}_i)=1-\frac{\Delta \rm p(\mathbf{r}_i)}{\rm max(\Delta \rm p(\mathbf{r}_i))},
\end{equation}
approaching 1 as $\Delta \rm p(\mathbf{r}_i) \rightarrow 0$ and approaching  0 when  $\Delta {\rm p}(\mathbf{r}_i) \rightarrow {\rm max}(\Delta \rm p(\mathbf{r}_i))$, ${\rm max}(\Delta \rm p(\mathbf{r}_i))$ being the maximum error estimated for the final sample.

We can associate to the $j$-th pixel of the Healpix map, whose position on the sky is $\mathbf{r_j}$, the parameter  $\rm p_\mathbf{j}$ and the error $\Delta \rm p_\mathbf{j}$ as follows:
%
%


Our spectral index and error maps are undefined in regions, mostly at high Galactic latitudes ($|b| > 30^\circ$), where the Galactic emissions are too low and the CCA cannot produce reliable estimates of their spectral parameters. For these regions we have adopted the mean values of the parameters found for the rest of the sky, using a smoothing function to soften edge effects.

\begin{figure*}
\begin{center}
\includegraphics[width=7cm]{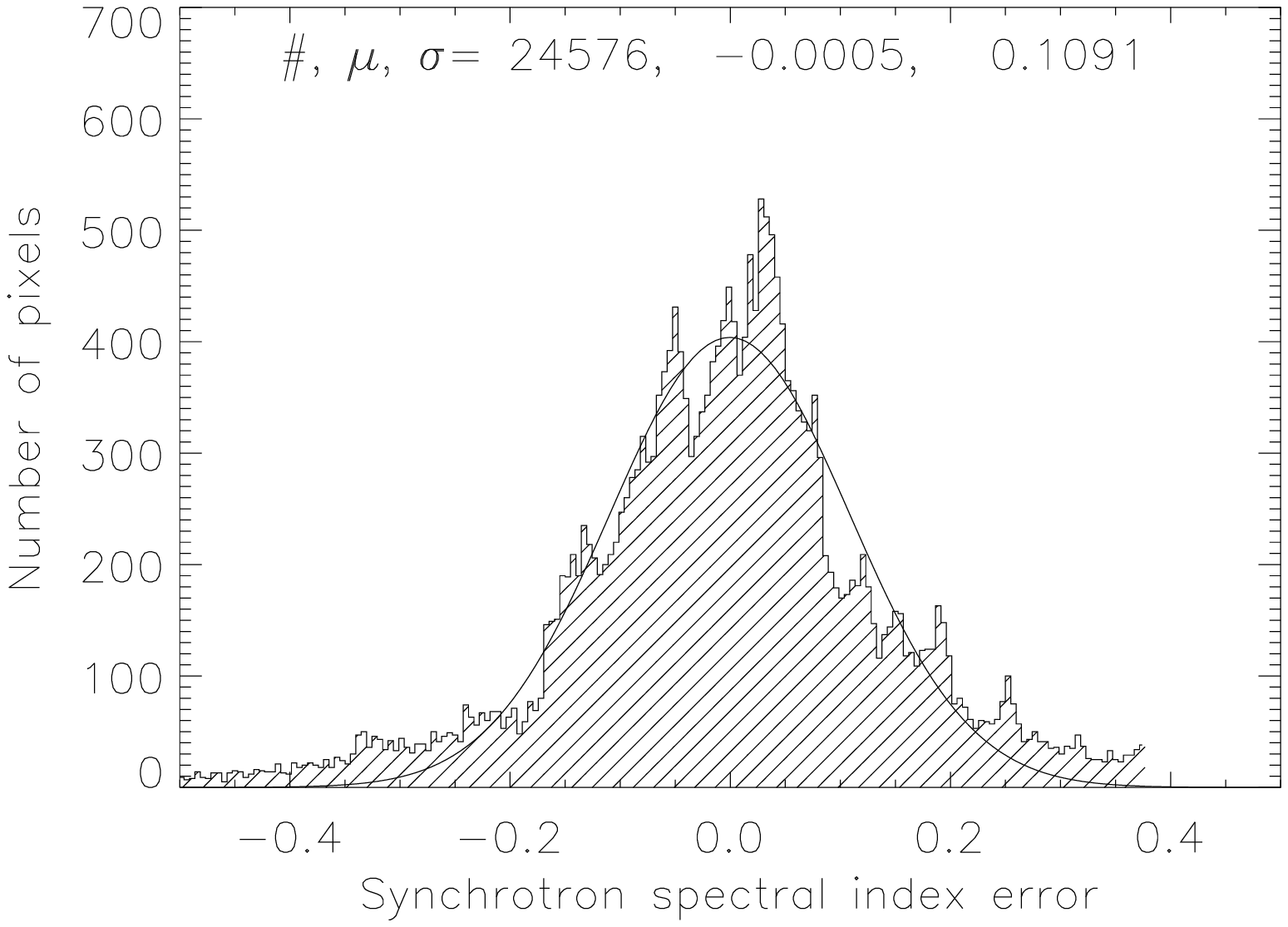}
\includegraphics[width=7cm]{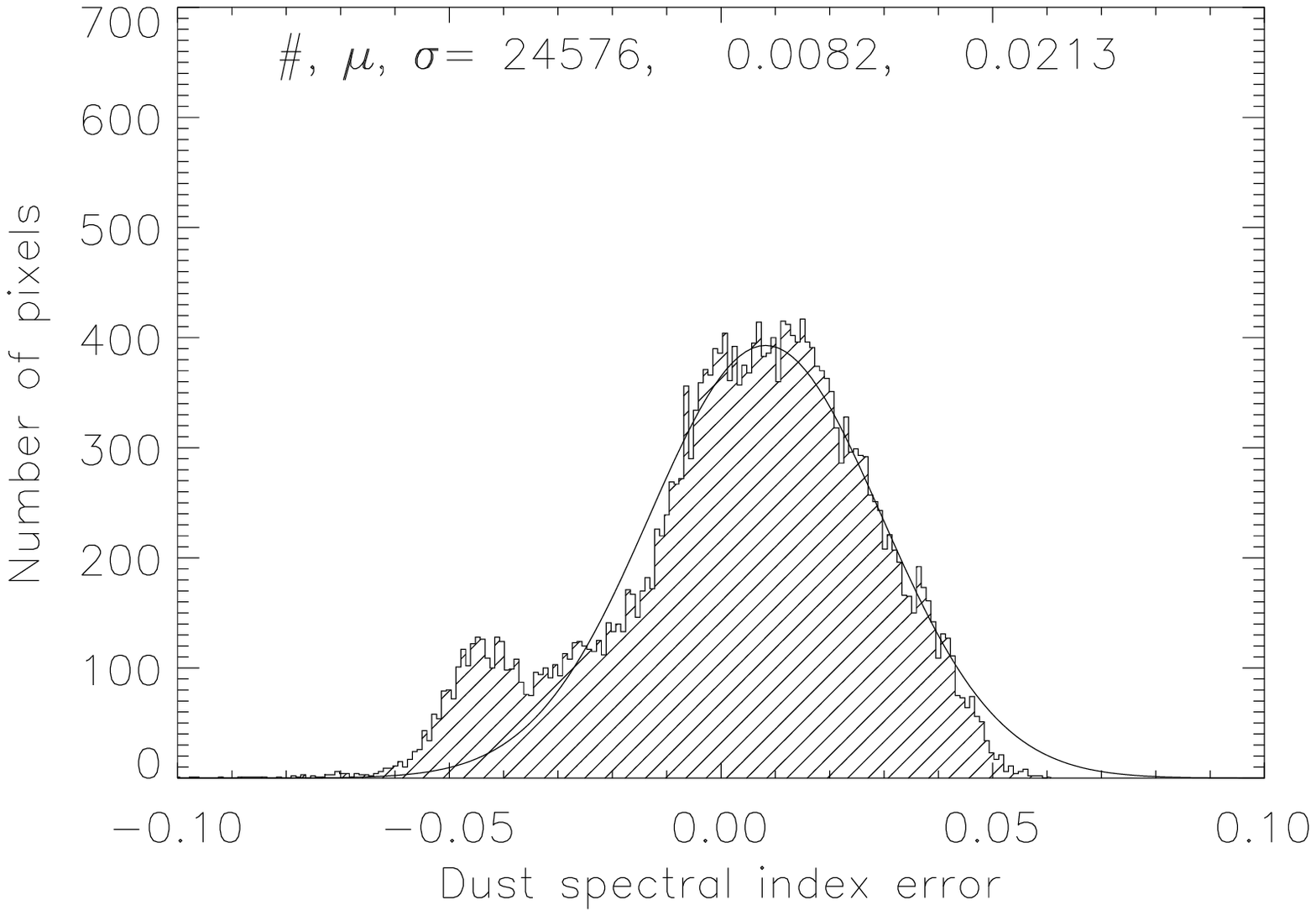}
\includegraphics[width=7cm]{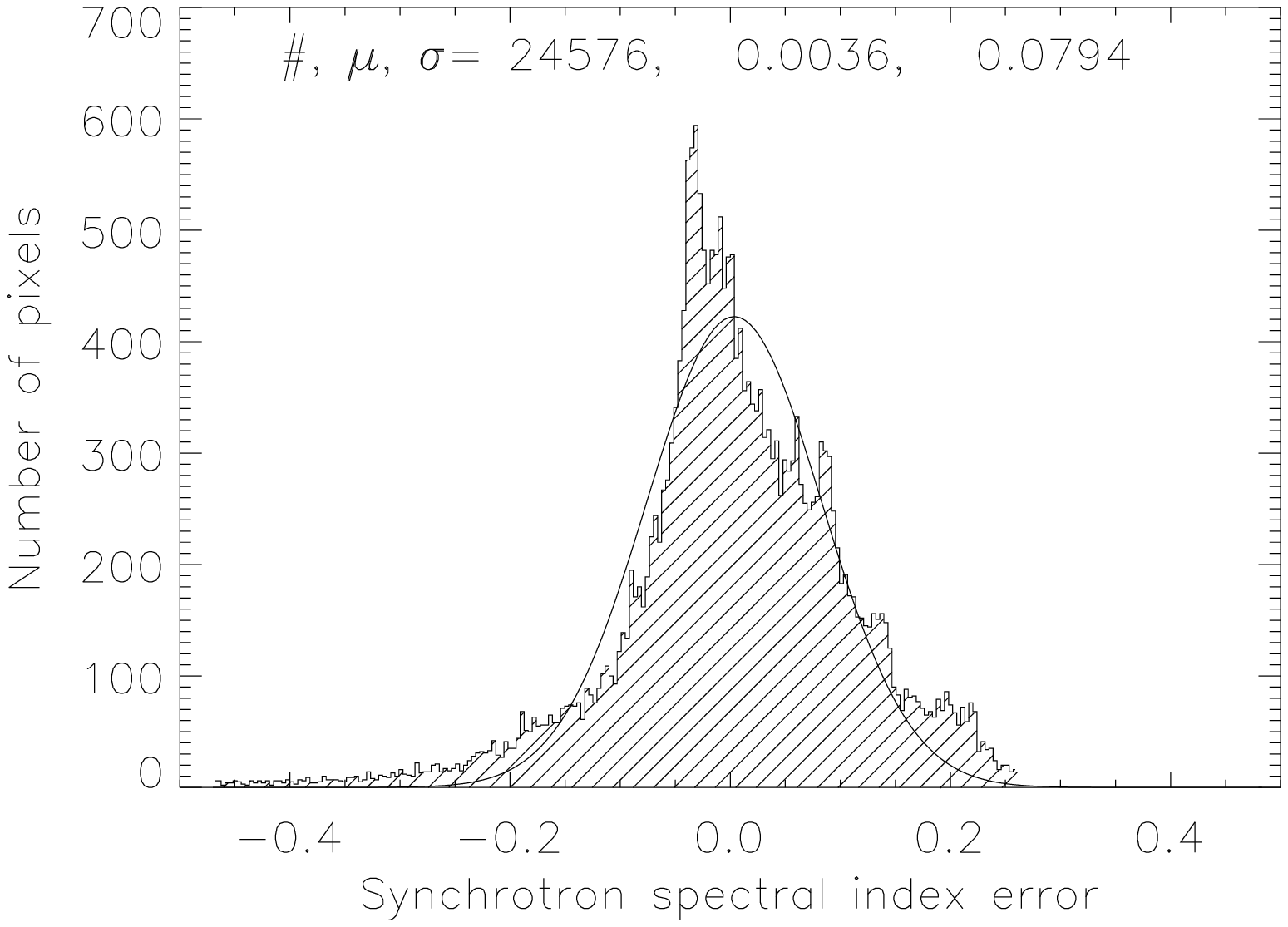}
\includegraphics[width=7cm]{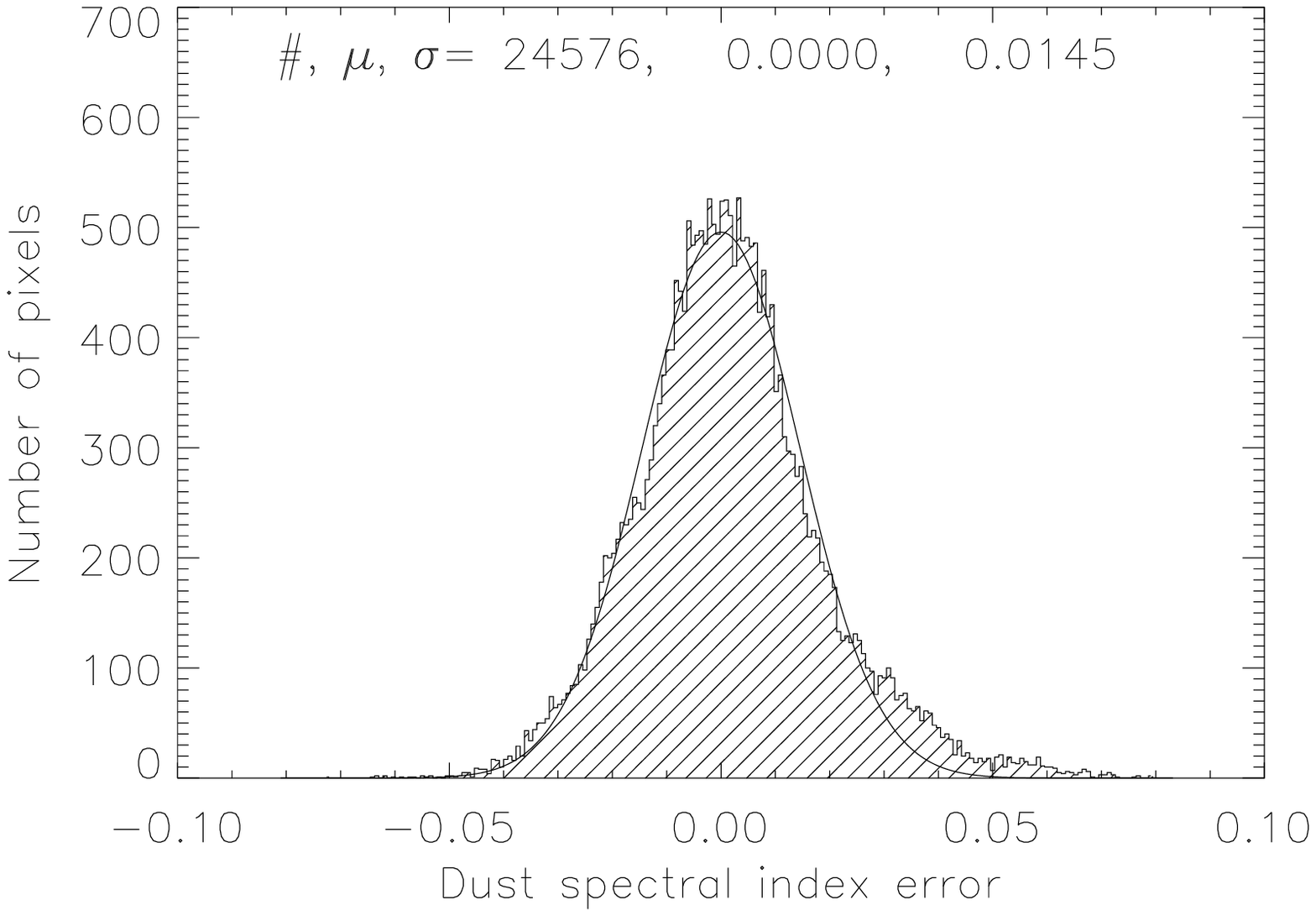}
\caption{Distribution of residual (true-estimated) synchrotron (left) and dust
  (right) spectral index maps in the latitude range   $-30^\circ<b<30^\circ$. The pixel size used is $\sim 0.84$ square degree (nside 64). Top: pixel domain CCA; bottom: harmonic domain CCA.} \label{fig:res_hist}
\end{center}
\end{figure*}

\subsection{Evaluation of the quality of the results}
To evaluate the quality of the maps of estimated spectral parameters, we
looked at the residual spectral index maps. To compute those maps, we
regridded the true input spectral index maps  to NSIDE=64, the resolution of
the estimated maps.  
In Fig.~\ref{fig:res_hist} we show the distributions of residuals (true minus
estimated spectral indices) with a Gaussian fit superimposed. The histograms
only contain the pixels in the latitude range $-30^\circ<b<30^\circ$ where the
spectral indices have been estimated. 
Overall, the results are encouraging. Again the harmonic domain CCA works
better than the pixel domain version for both dust and synchrotron spectral
parameters. Both approaches give negligibly small mean offsets from the true
values, but the harmonic CCA has smaller dispersions of the residuals
($\sigma_s \simeq 0.08$, $\sigma_d \simeq 0.01$) than the pixel domain CCA
($\sigma_s \simeq 0.11$, $\sigma_d \simeq 0.02$).

%% file: sec7.tex
Having shown that the harmonic domain CCA is superior to the pixel domain
version, we have used its estimates of synchrotron and dust spectral indices
to build the spatially-varying mixing matrix $\mathbf{H}$ and the
reconstruction matrix $\mathbf{W}$ given by eq.~(\ref{gls}). To recover each
emission component we have applied the reconstruction matrix [see
eq.~(\ref{recon})] to the simulated data for {\sc Planck} channels in the
frequency range from 70 to 217 GHz, where the CMB is the least contaminated by
foreground emissions. On the other hand, just for the same reason, this
frequency range is not optimal for the recovery of foreground emissions. All
the maps were smoothed to the resolution of the 70 GHz channel (14 arcmin
FWHM). We also performed a run at a 60 arcmin resolution to reduce the
contribution of instrumental noise, so that we can better appreciate the
effect of component separation errors on the reconstructed maps. Besides the
component maps [eq.~(\ref{recon})], we also computed the variance due to
instrumental noise [eq.~(\ref{covnoise})] and to residual foreground
contamination. The latter is estimated by propagating the errors on the mixing
matrix parameters to the separated components as descibed in
\S\, \ref{err_recon_components}. 
In particular we assume that the spatial correlation of separation errors in
the spectral index in each pixel are the same as those in the spectral index
map itself. Thus, in order to evaluate and propagate the error in the outputs,
we construct Monte Carlo variations of the estimated spectral index maps. In
doing this, we further simplify our treatment by only taking into account the
two-point correlation function of the map. 

Therefore, if $\beta_s$ and $\beta_d$ are the maps of the estimated
synchrotron and dust spectral indices respectively, and  $err\beta_s$,
$err\beta_d$  the corresponding estimated error maps, for the i-th run we obtain the perturbed spectral index maps $\beta_s^i$ and $\beta_d^i$ as follows:
\begin{eqnarray}
\label{perturb}
{\rm \beta_s^i}&=&{\rm \beta_s}+\Delta_{s}\cdot {\rm err\beta_s}\\
{\rm \beta_d^i}&=&{\rm \beta_d}+\Delta_{d}\cdot {\rm err\beta_d},\nonumber
\end{eqnarray}
where $\Delta_{s}$ and $\Delta_{d}$ are two maps with zero mean and unit
rms synthesized from the 2-point correlation function extracted from the
corresponding estimated spectral index map.
\begin{figure*}
\begin{center}
\includegraphics[angle=0,width=7.8cm,height=5.2cm]{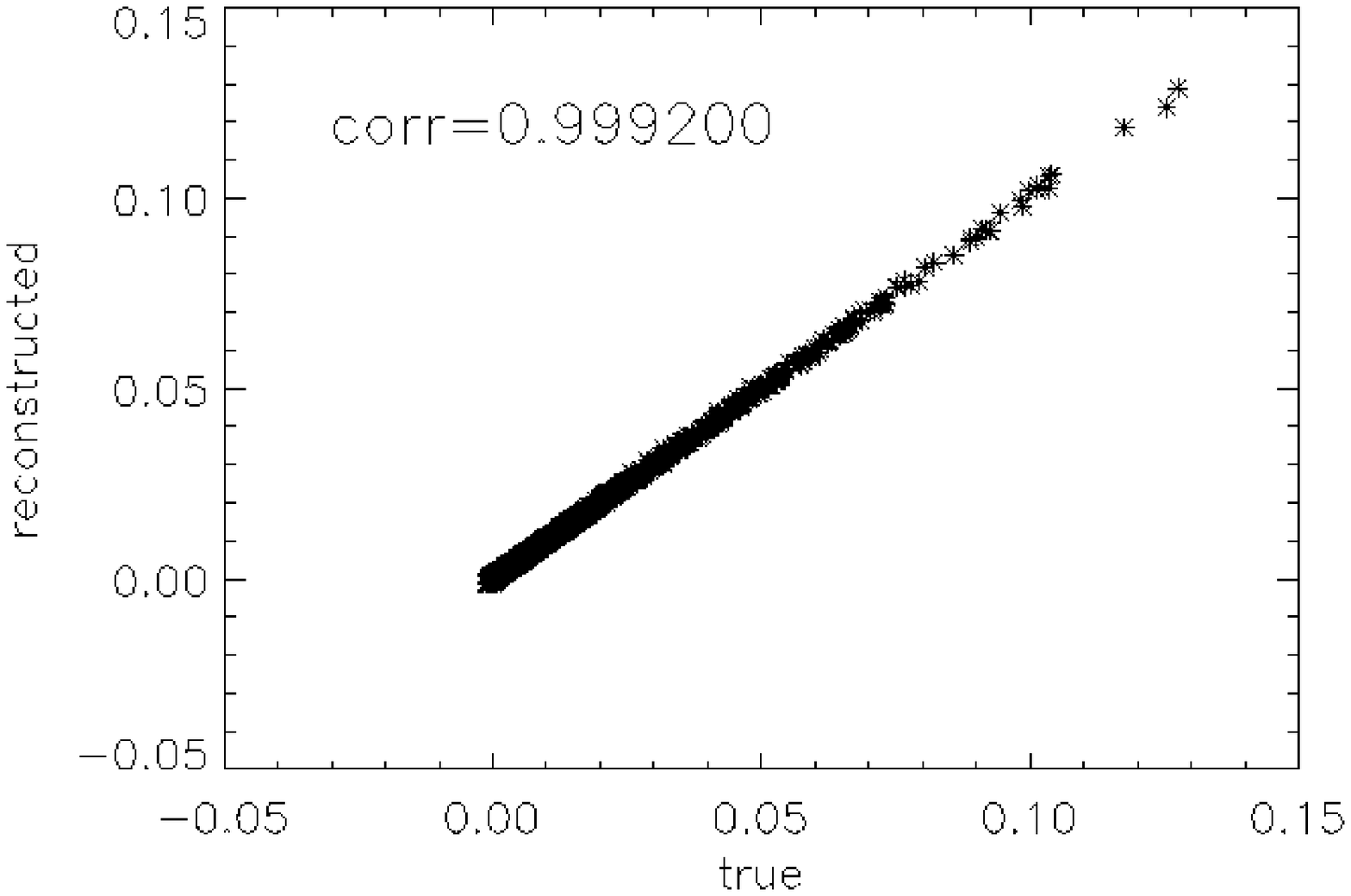}
\includegraphics[angle=0,width=7.8cm,height=5.2cm]{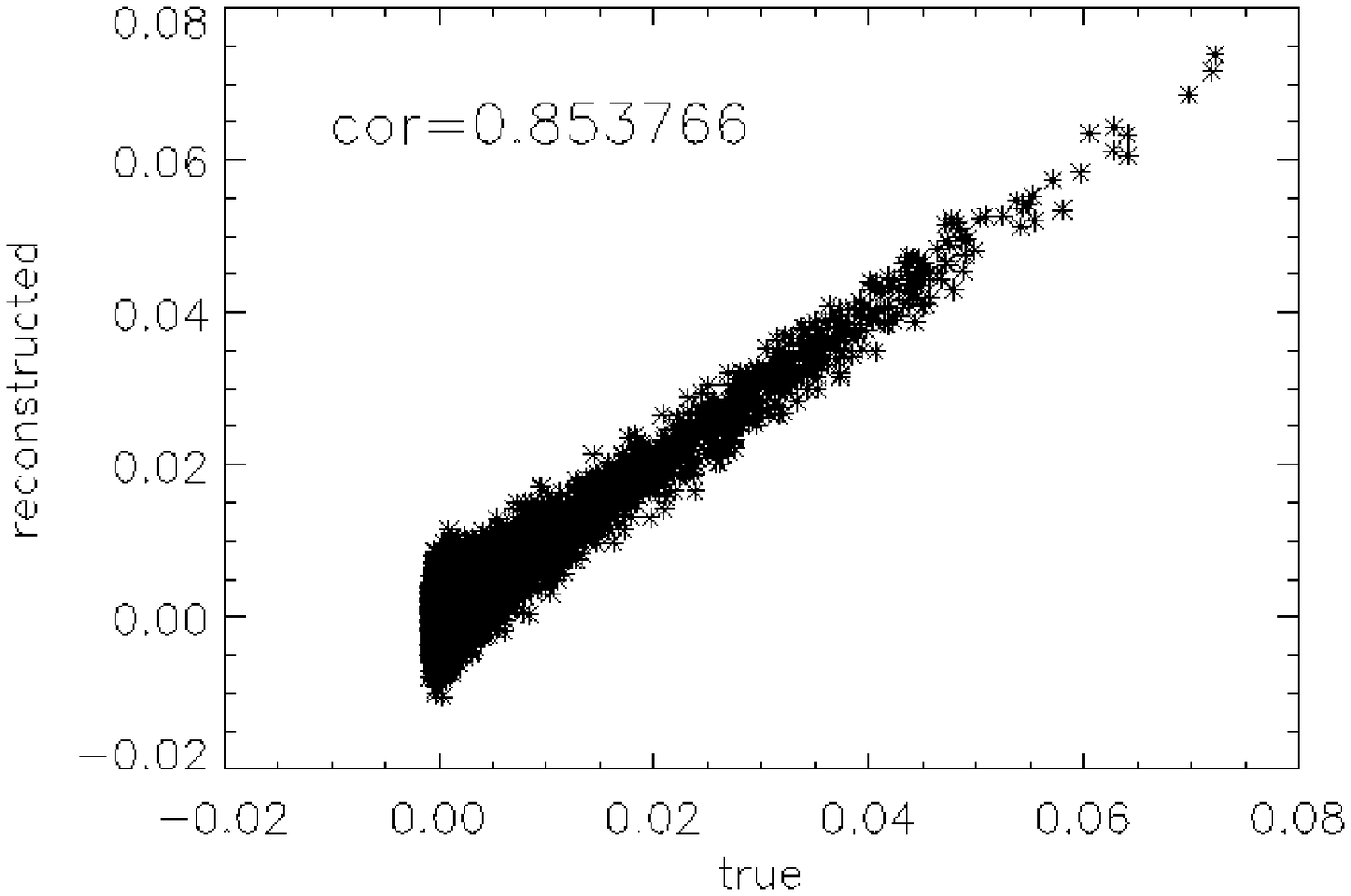}
\includegraphics[angle=0,width=7.8cm,height=5.2cm]{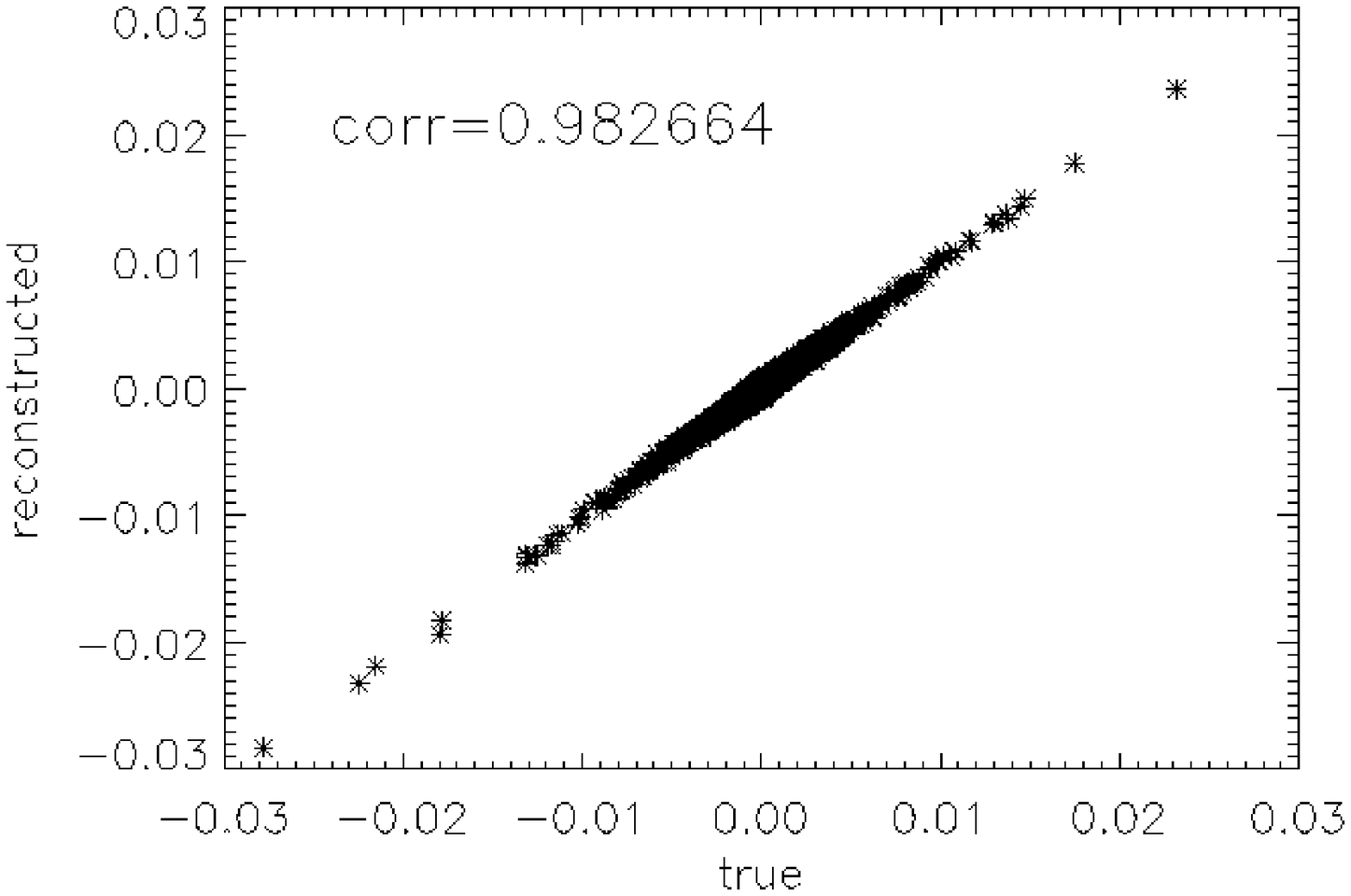}
\includegraphics[angle=0,width=7.4cm,height=5.2cm]{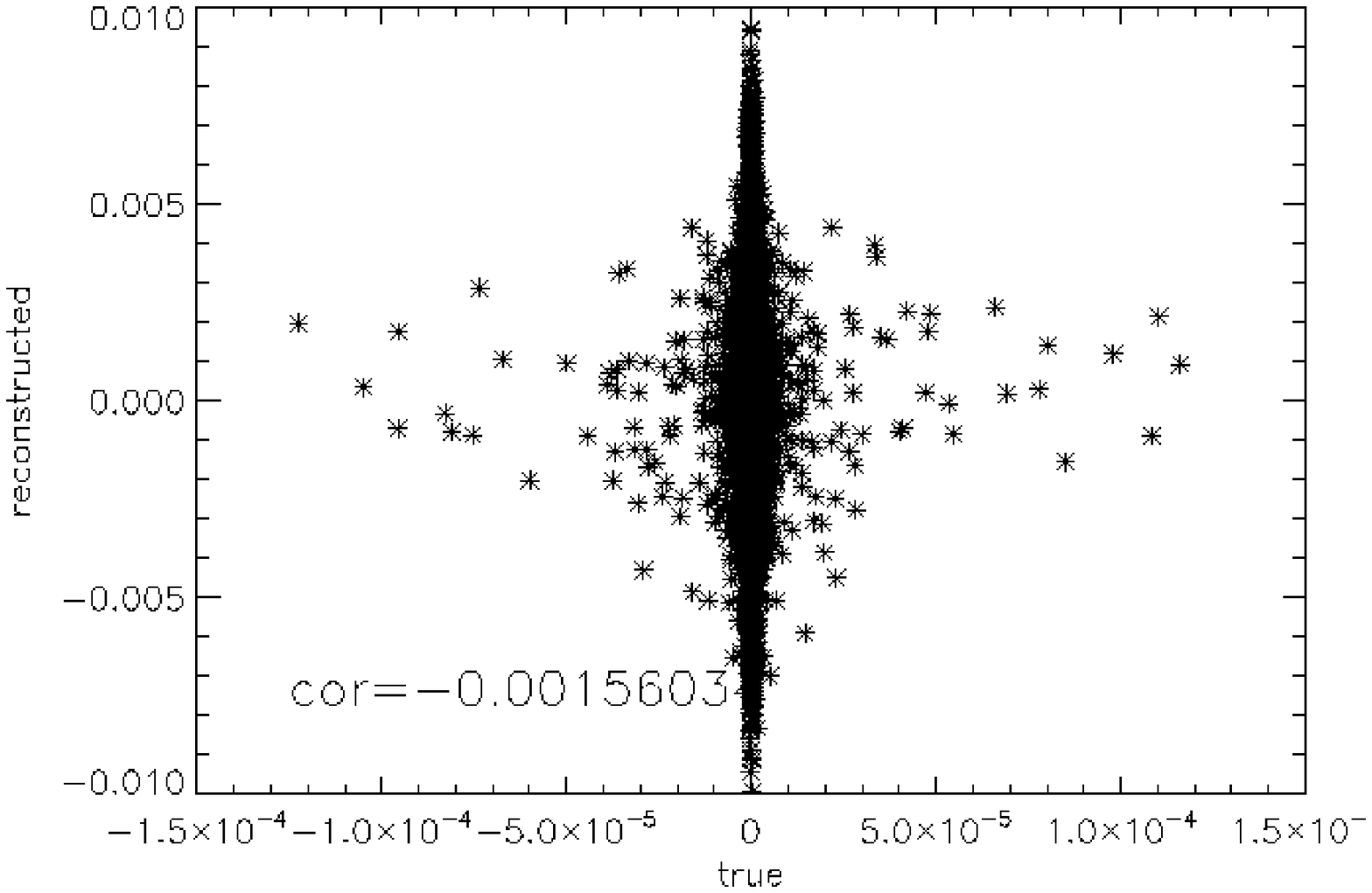}
\caption{Correlations between the true and the reconstructed $Q$ and $U$ maps
  of Galactic components for $\simeq 1^\circ$ pixels. Upper left panel: dust
  $Q$  (correlation coefficient $\simeq 0.99$); upper right panel: synchrotron
  $Q$ (correlation coefficient $\simeq 0.85$); lower left panel: dust $U$
  (correlation coefficient $\simeq 0.98$ ). The synchrotron $U$ (lower right panel) was not detectable. \label{fig:scatterplot}}
\end{center}
\end{figure*}
%


For each set of fake spectral index maps obtained through eqs.~(\ref{perturb}) we performed the source reconstruction by GLS, thus obtaining as many sets of output components.
We  computed the variance due to separation for a certain component as the variance of all the results for that component.
\begin{figure}
\begin{center}
\includegraphics[width=6cm,angle=90]{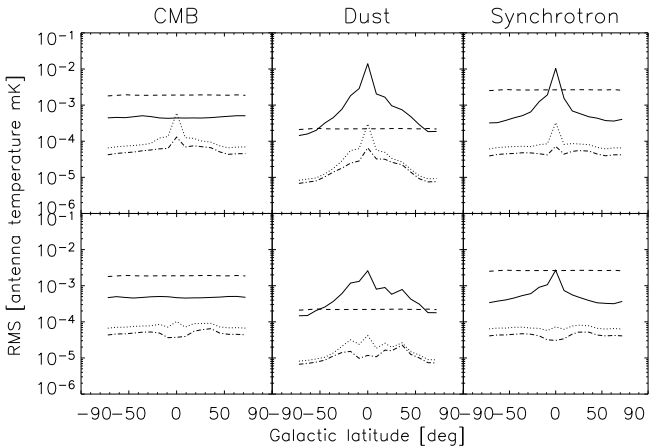}
\caption{True rms fluctuations of input polarized components at the reference
  frequency of 70 GHz as a function of Galactic latitude for a $60'$
  resolution (solid line) compared with estimated error due to instrumental
  noise (dashed line) and estimated error due to residual contamination for
  the marginal distribution (dotted line) and spatial redundancy
  (dashed-dotted line) error estimation methods. Upper panels: Q Stokes
  parameter; lower panels: U Stokes parameter.\label{fig:comp_var} }
\end{center}
\end{figure}

In Fig.~\ref{fig:comp_var} we show for all the components reconstructed with $60'$ resolution the rms of the true input component as a function of Galactic latitude, compared to the $\sigma$ due to noise and imperfect separation, computed from the variances output by our method. Even at this low resolution instrumental noise dominates except close to the Galactic plane where the S/N ratio is $\sim 10$. The error estimation yielded by the marginal distribution method is more conservative than the one obtained with spatial redundancy method; the two estimates typically differ by 30\%. 

\begin{figure*}
\begin{center}
\includegraphics[angle=0,width=7.5cm,height=5.2cm]{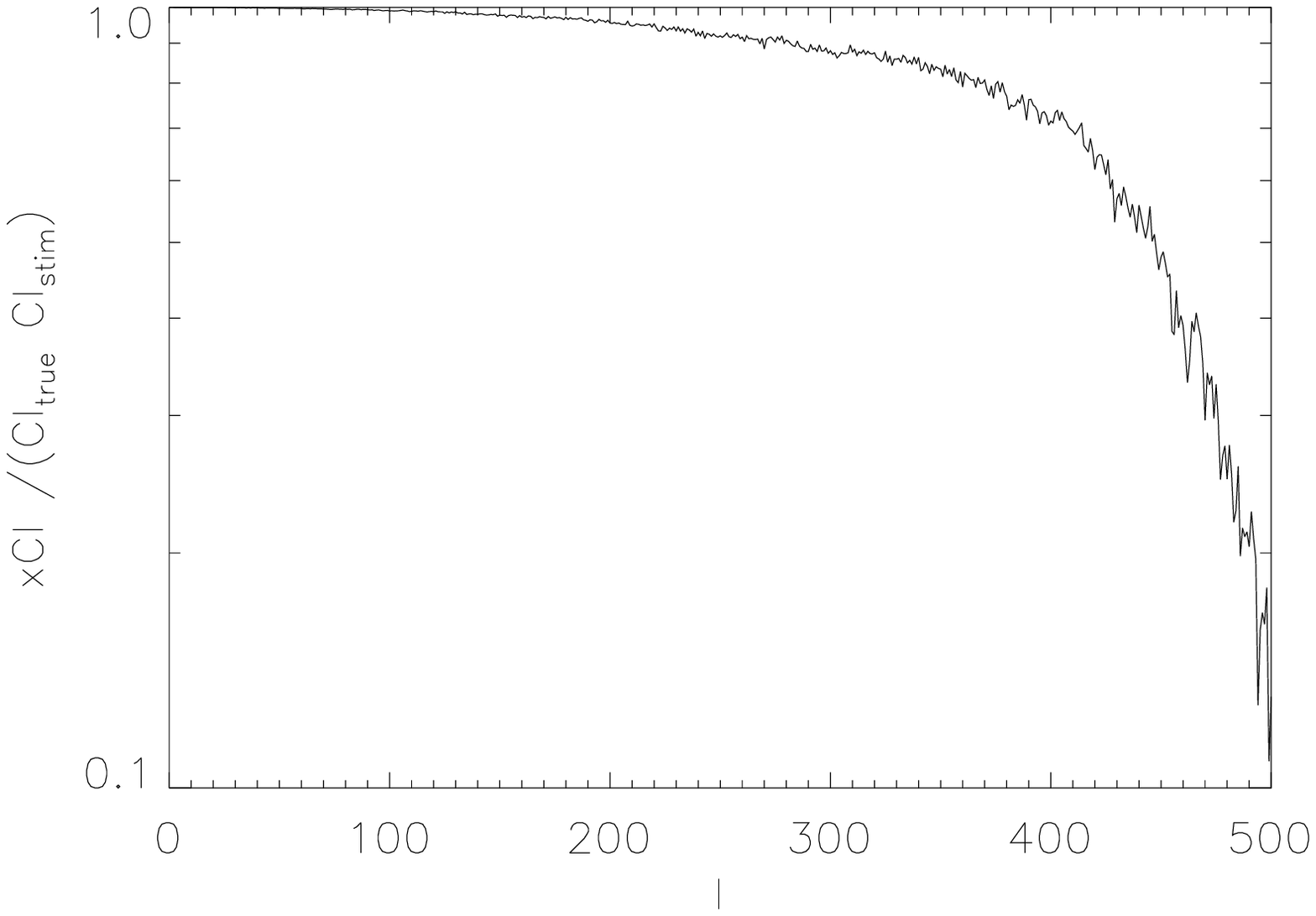}
\includegraphics[angle=0,width=7.5cm,height=5.2cm]{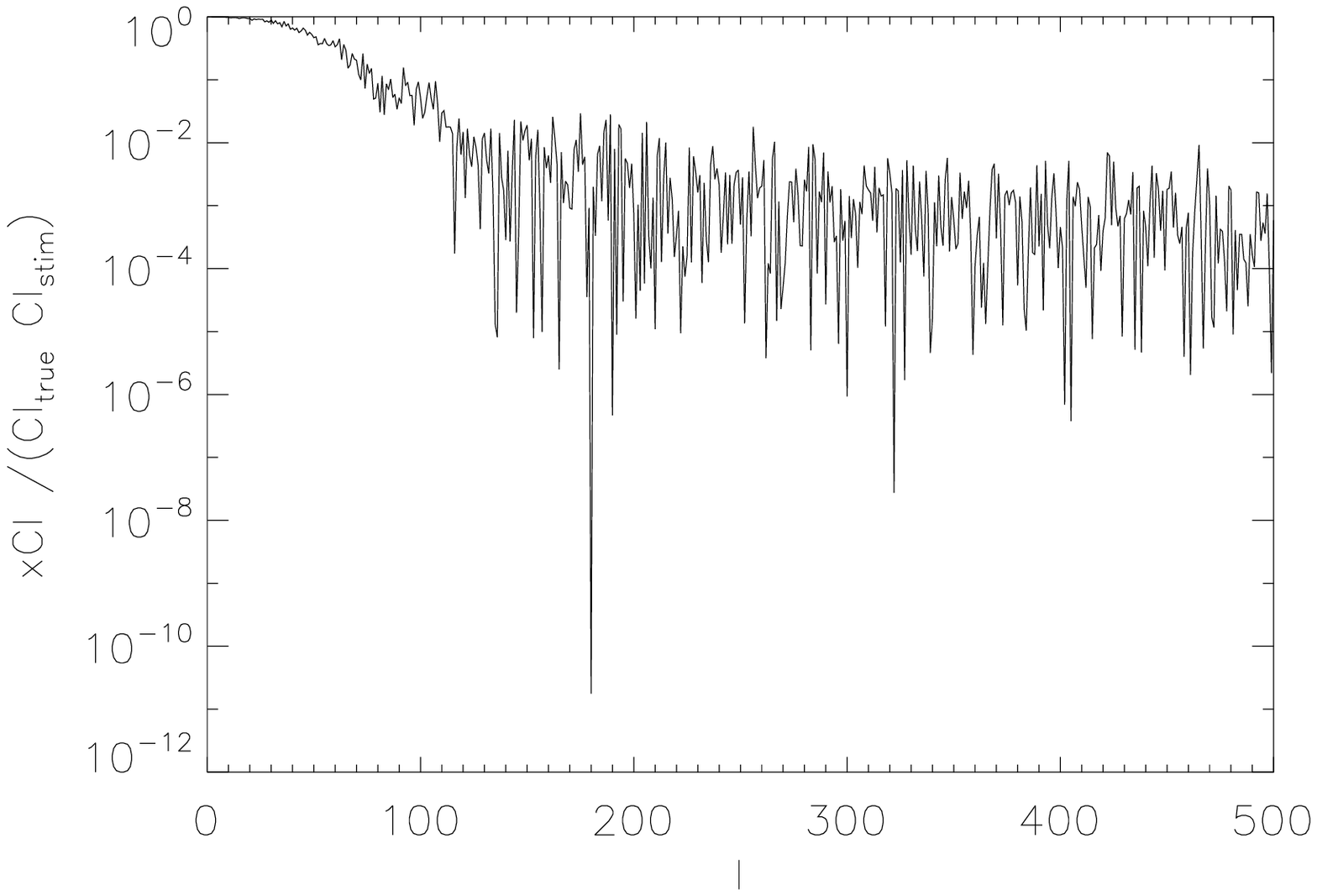}
\caption{Cross spectra [eq.~(\protect\ref{eq:cross})] between the true and the reconstructed  component for dust (left) and synchrotron emission (right).\label{fig:cross_spettrifore}}
\end{center}
\end{figure*}

\begin{figure*}
\begin{center}
\includegraphics[width=4.5cm,angle=90]{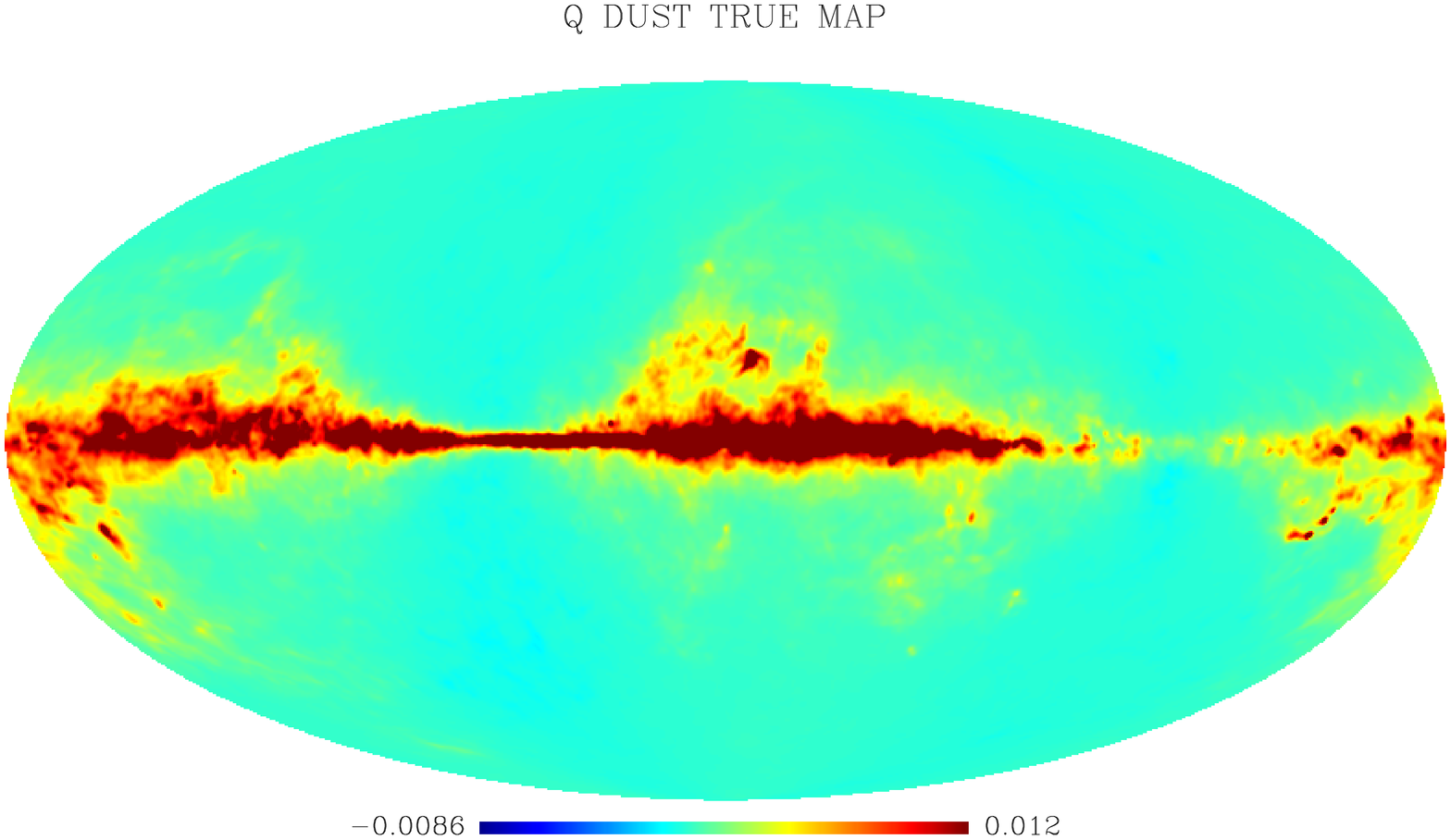}
\includegraphics[width=4.5cm,angle=90]{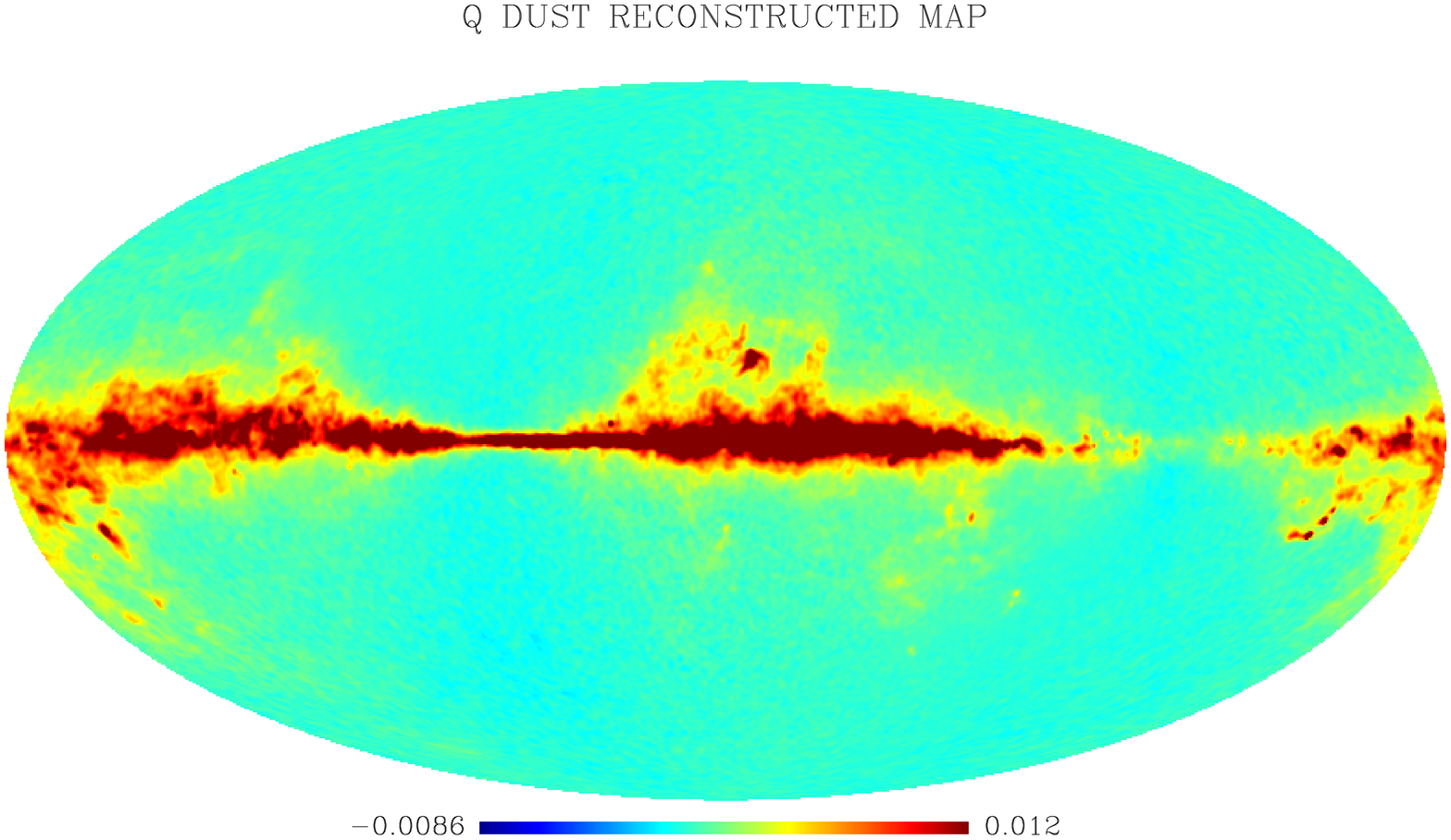}
\includegraphics[width=4.5cm,angle=90]{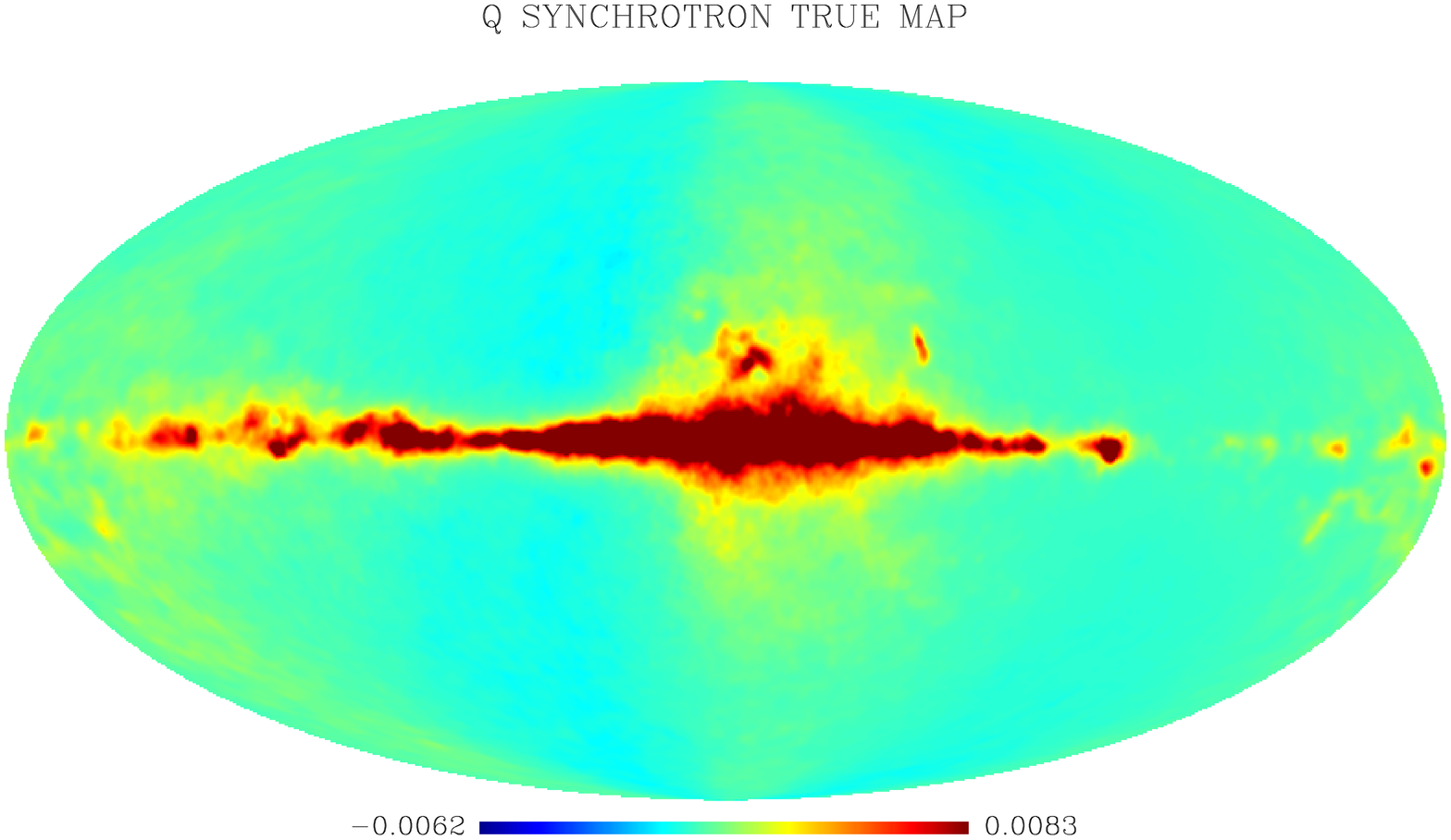}
\includegraphics[width=4.5cm,angle=90]{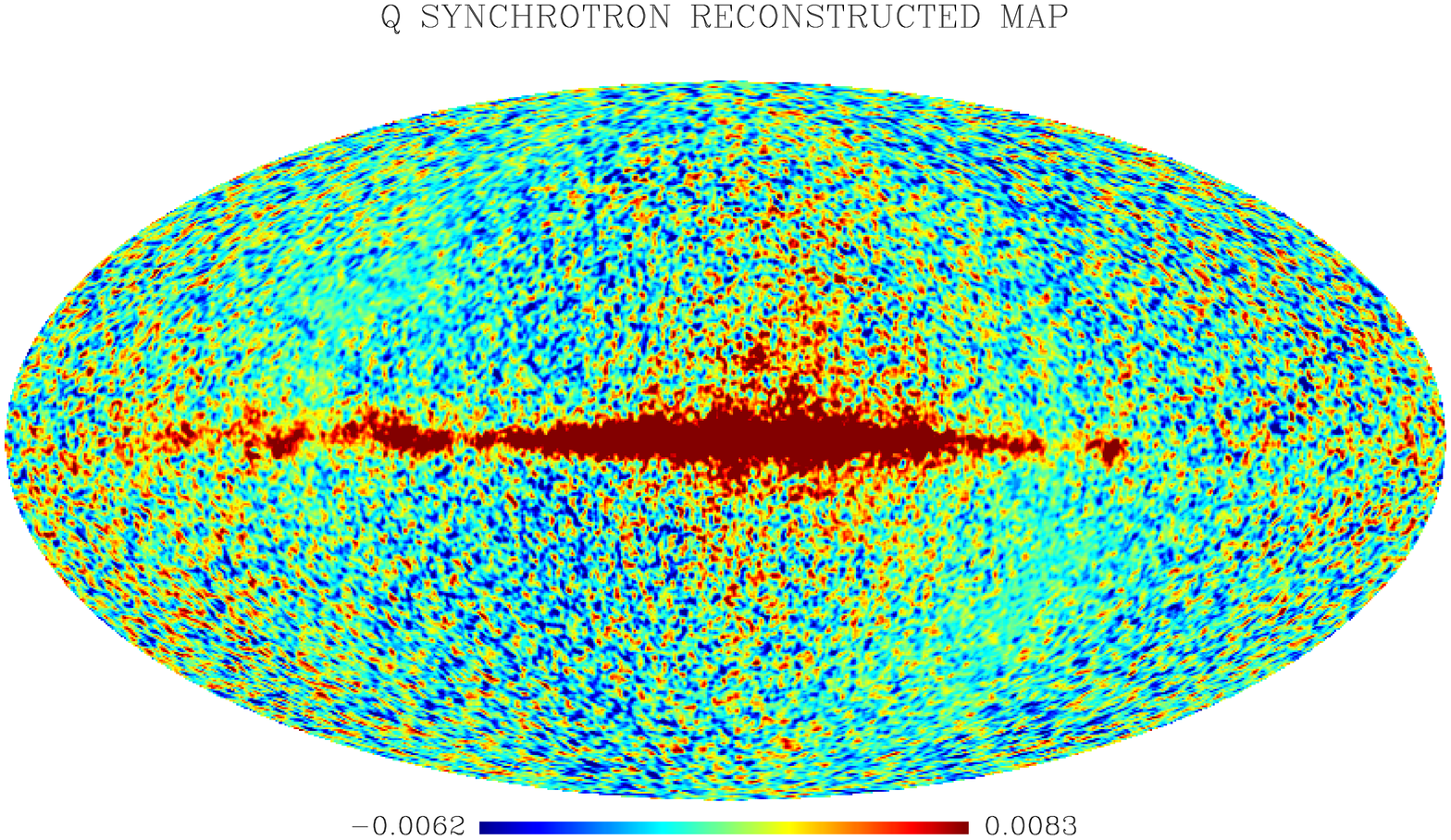}
\caption{True and  reconstructed maps for dust (top) and synchrotron emission (bottom).\label{fig:mappefore}}
\end{center}
\end{figure*}
\begin{figure*}
\begin{center}
\includegraphics[width=10cm,angle=90]{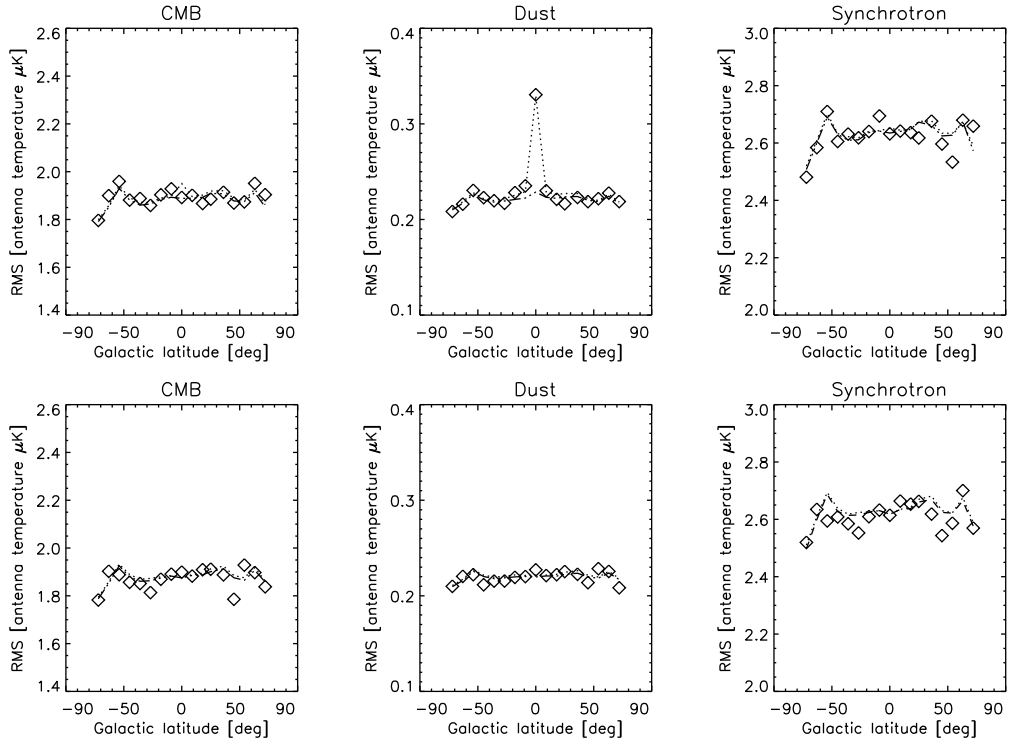}
\caption{Comparison of true (diamonds) and estimated rms errors as a function of Galactic latitude at 70 GHz for $Q$ (upper panels) and $U$ (lower panels) maps. Dotted lines: marginal distribution method (\S\,\protect\ref{sec:marginal}); dot-dashed lines: spatial redundancy method (\S\,\protect\ref{sec:redundancy}).\label{fig:merit_var}}
\end{center}
\end{figure*}
\begin{figure*}
\begin{center}
\includegraphics[width=10cm,angle=90]{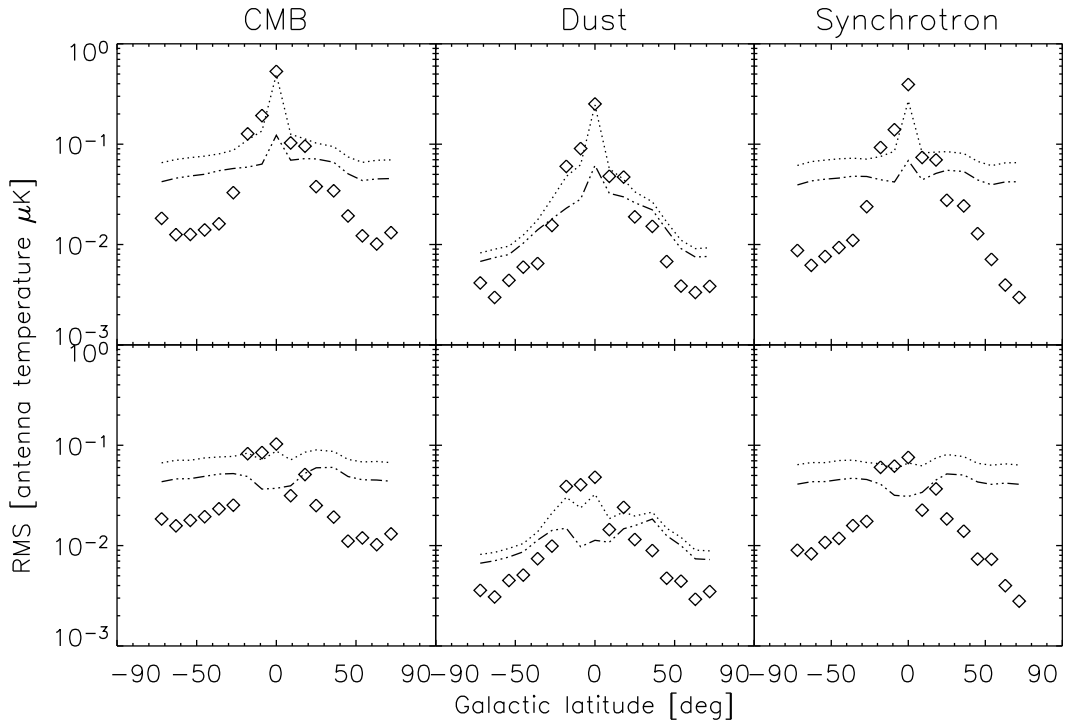}
\caption{Noiseless case: comparison of true (diamonds) and estimated rms errors as a function of Galactic latitude at 70 GHz for $Q$ (upper panels) and $U$ (lower panels) maps. Dotted lines: marginal distribution method (\S\,\protect\ref{sec:marginal}); three dots-dashed lines: spatial redundancy method (\S\,\protect\ref{sec:redundancy}). At high Galactic latitudes, where both our methods overestimate the component separation errors, such errors are irrelevant compared to uncertainties due to instrumental noise. \label{fig:merit_var_noiseless}}
\end{center}
\end{figure*}
\begin{figure*}
\begin{center}
\includegraphics[width=5cm,angle=90]{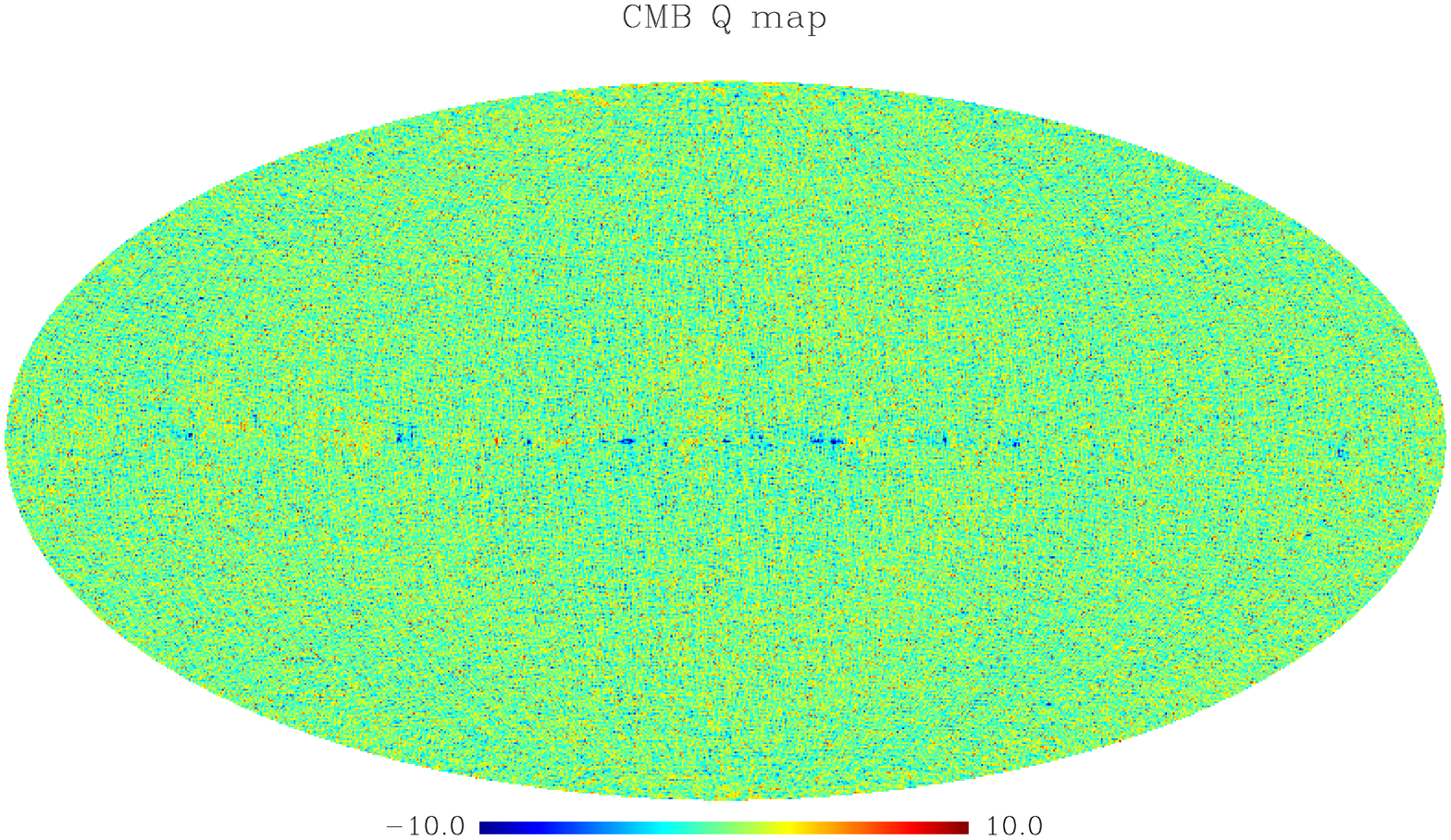}
\includegraphics[width=5cm,angle=90]{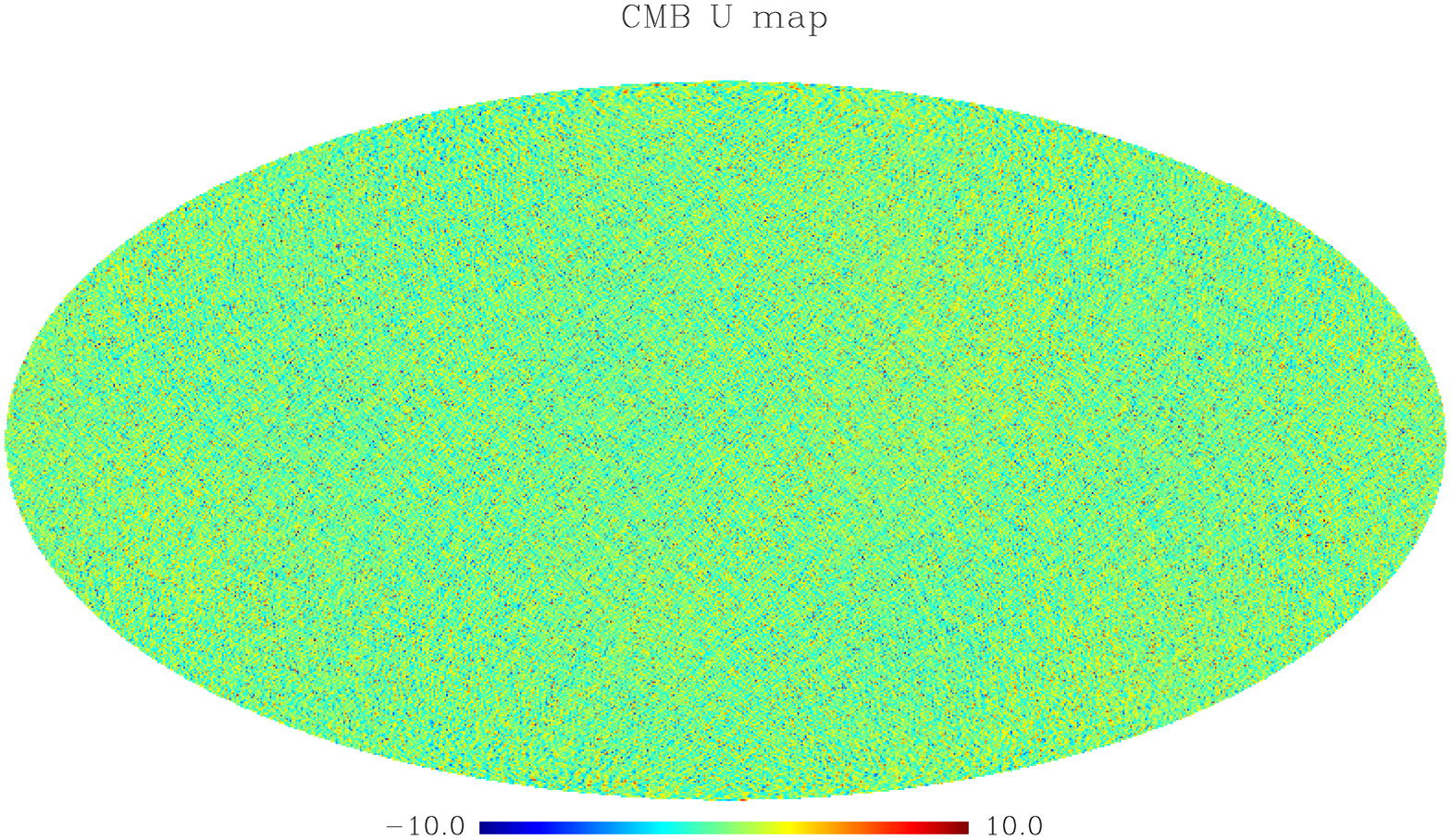}
\caption{Full-sky $Q$ (left) and $U$ (right) CMB maps obtained by CCA for the noiseless case.\label{fig:q_u_maps}}
\end{center}
\end{figure*}

\subsection{Quality of the reconstructed component maps}
%
In Fig.~\ref{fig:scatterplot} we compare the recovered $Q$ and $U$ maps of
synchrotron and dust emission with NSIDE=64 (resolution $\simeq 1^\circ$). The
agreement is almost perfect for dust (correlation coefficient of 0.98--0.99),
and very good for the synchrotron $Q$  (correlation coefficient of 0.85). The
synchrotron $U$ map could not be reconstructed because of the very low S/N
ratio (see Fig.~\ref{fig:comp_var}) due to the choice of frequency channels,
that left out the low frequency ones (30 and 44 GHz, and WMAP 23 GHz) where
the synchrotron is much stronger. 

%
As another figure of merit, we computed the normalized correlation multipole by multipole between the true and reconstructed map:
%
\begin{equation}\label{eq:cross}
\frac {\langle C_{\ell}^{\rm cross}\rangle^2} {\langle C_{\ell}^{\rm true}\rangle \langle C_{\ell}^{\rm stim}\rangle },
\end{equation}
where $C_{\ell}^{\rm true}$ and $C_{\ell}^{\rm stim}$ are the auto-spectra of the true and the reconstructed component, and $C_{\ell}^{\rm cross}$ the cross-spectrum between the two. Figure~\ref{fig:cross_spettrifore} shows that, for dust, the cross-correlation is close to unity on all the relevant scales. The same is true for synchrotron, on scales larger than a few degrees; on smaller scales the correlation drops because the instrumental noise dominates.
In  Fig.~\ref{fig:mappefore} we show the comparison between the true and the reconstructed Q synchrotron and dust emission maps.


\subsection{Assessment of the component separation error estimations}
As outlined above, our method outputs two rms error maps, one for instrumental
noise and the other for component separation errors. The actual error map for
each component is simply computed as the residual map, i.e. the difference
between the reconstructed map of a component and the true one at the same
resolution. This residual map contains both noise and component separation
errors, and thus has to be compared to the global estimated error map,
obtained summing the variances of the two estimated contributions. 

In Fig.~\ref{fig:merit_var} we show the standard deviations  of the residuals as  function of
Galactic latitude (diamonds) compared to the total error
yielded by both our error estimation methods for the 60 arcmin
resolution  (lines). The results are generally very good, as the total error is
correctly predicted. On the other hand, since we are generally noise
dominated, this figure cannot tell much about the goodness of our component
separation error estimates, except for dust close to the Galactic plane, where
the error estimate turns out to be very successful. We note however that, as
we are using a linear estimator [eq.~(\ref{recon})] to reconstruct the
components, our final results can be decomposed into a noiseless term, only
affected by component separation errors, plus a noise term: 
\begin{equation}
\label{eq:quarantadue}
\mathbf{\hat{s}}=\mathbf{Wy}=\mathbf{W[Hs]}+\mathbf{Wn}.
\end{equation}
Here $\mathbf{Hs}$ is a noiseless dataset, obtained exactly as described in
\S\,\ref{sec:dataset} but without adding noise. By combining the noiseless
dataset with  \emph{the same reconstruction matrix estimated in the noisy
  case}, we obtain a set of reconstructed components having the same component
separation errors as before, but without noise. This allows us to test the
quality of component separation error estimates. The results, shown in
Fig.~\ref{fig:merit_var_noiseless}, are very encouraging. Even if component
separation errors are in general highly subdominant, the marginal distribution
error estimation method is able to correctly estimate them at low and
intermediate Galactic latitudes (at high Galactic latitudes the component
separation errors are overestimated but are anyway irrelevant compared to
errors due to noise). On the other hand  the spatial redundancy error
estimation method is occasionally underestimating the true errors at low
latitude. 
In Fig.~\ref{fig:q_u_maps} we show, as an example, the CMB $Q$ and $U$
reconstructed maps in the noiseless case. A visual inspection does not reveal
the presence of residual Galactic contamination except for a tiny strip on the
Galactic plane. 

%% file: sec8.tex
To assess the quality of the Stokes $Q$ and $U$ CMB maps obtained with the CCA
component separation we compare their estimated angular power spectra (APS) to
the input model used to generate the simulation. In doing so we propagate to
the power spectra the component separation errors described above. We employ
the ROMAster code, a pseudo-$C_\ell$ estimator based on
MASTER approach ~\citep{master} and extended to cross-power
spectra~\citep{polenta2005} and polarization \citep[see e.g.][for a similar
formalism]{kogut}. It is well known that the pseudo $C_\ell$ approach to the
CMB power spectrum estimation is sub-optimal for the lowest multipoles where
other techniques are more appropriate \cite[see, e.g.,][]{bolpol}. However, a
pseudo-$C_\ell$ estimator is enough for our purpose of assessing the quality
of the reconstructed CMB polarization maps in the presence of noise and
component separation errors. 

\begin{figure}
\begin{center}
\includegraphics[width=5cm,angle=90,keepaspectratio]{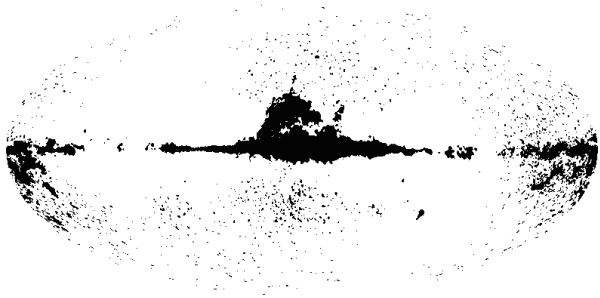}
\caption{Mask used for the computation of the EE power spectrum, excluding 9\% of the sky on the basis of the estimated variance due to component separation on the CMB Q and U maps.\label{fig:EE_mask}}
\end{center}
\end{figure}

We exclude from the analysis the regions that are most contaminated by
residual foreground contributions as estimated in the previous section. For
this purpose we build a mask based on our reconstruction errors, flagging all
pixels where the sum of the variance errors on the CMB $Q$ and $U$ maps is
greater than  the mean value of the same quantity across the whole map. 
The resulting mask is shown in Fig.~\ref{fig:EE_mask} and excludes less than
10\% of the sky. 

Having only one final  map per astrophysical component, we do not rely here on
a cross spectrum analysis but on an auto-spectrum approach. This is rather
general, and achieves a lower final noise variance than the cross spectrum
approach \citep[see, e.g.,][]{polenta2005}. The drawback is that we need to
model and subtract a noise bias in the data. To this extent, we computed the
noise bias on the CMB EE power spectrum by means of 1000 simulated noise
maps. To obtain each of them, we simulated one noise map for each channel
included in the reconstruction of the CMB (70, 100, 143 and 217 GHz),
equalized the resolution of all channels to $14'$, and combined them with the
reconstruction matrix  $\mathbf{W}$ as described in the previous section. 
ROMAster uses these Monte Carlo data to subtract the noise bias, as well as to
estimate errors  on the APS due to noise by computing the empirical variance
of the realization. 

To compute the error bars due to residual foreground contamination, we
produced a further set of 100 CMB maps by perturbing the input spectral index
maps as described in the previous section. For each of them we repeated the
computation of the power spectrum and corrected for the noise bias, relying
for the latter purpose on a smaller ($\sim 10$) set of noise-only maps. The
noise bias has been estimated each time with the reconstruction matrix used to
obtain the corresponding CMB map. Even if quite computationally demanding,
this procedure is needed because we are in the noise-dominated regime, and a
small error in the noise bias can substantially affect the estimation of the
polarized power spectrum. Once we got our 100 unbiased CMB power spectra, we
finally computed the errors due to component separation as the standard
deviation of the sample for each considered multipole bin. 

\begin{figure}
\begin{center}
\includegraphics[width=8cm,keepaspectratio]{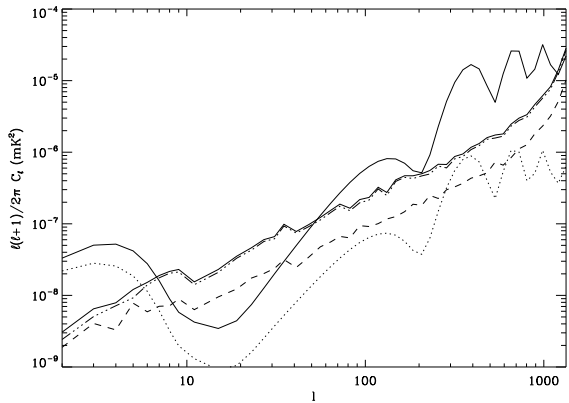}
\caption{EE power spectrum of the input model (solid) compared to the rms
  errors due to noise (dot-dashed line), separation (dashed line) and total
  (solid line). The dotted line represents  the associated cosmic variance
  error.\label{fig:errori_percent}}
\end{center}
\end{figure}

\begin{figure}
\begin{center}
\includegraphics[width=8cm,keepaspectratio]{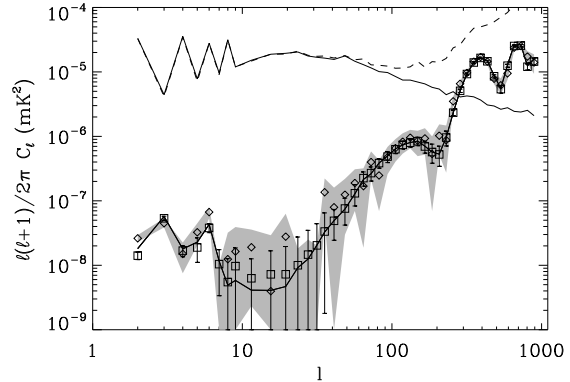}
\caption{ Recovered EE power spectrum in the realistic Planck case (diamond)
  and 1\,$\sigma$ uncertainties (shaded area);  recovered EE power spectrum in
  the noiseless case (squares) with 1\,$\sigma$ component separation error
  bars; EE input CMB map (gray solid line). On top of the plot we also show,
  for comparison, the total Galactic emission of our simulation at 70 GHz
  (solid line); the dashed line includes the effect of the 70 GHz {\sc Planck}
  noise. \label{fig:ee_sigonly}}
\end{center}
\end{figure}

In Fig.~\ref{fig:errori_percent} we show the noise and the component
separation error bars compared to the EE power spectrum of the fiducial
model. As we can see, the noise contribution is dominating even on the
smallest multipoles over the component separation error, which is at most a
small correction to the error budget. 
In Fig.~\ref{fig:ee_sigonly} we show the EE power spectrum estimated in a
realistic {\sc Planck} case (diamonds); the 1$\sigma$ errors are shown by the
shaded area. The results for the noiseless case (squares with the component
separation error bars) show that the accuracy of our estimation of the power
spectrum is not limited by component separation but rather by the effect of
the noise. 
The overall quality of the recovered spectrum is impressive, especially when
compared to the total Galactic emission at 70 GHz, shown on top of the plot. 
Although a discussion on the detectability of CMB B-modes is beyond the scope
of this paper \citep[see][for recent analyses]{bet2009,efst1}, we briefly
address here the potential of the CCA in this respect. An interesting point,
raised by \cite{efst2}, is that some foreground subtraction methods, such as
the ILC, are not well-suited for applications to B-mode detection, as they
introduce an offset caused by the dominant E-mode polarization pattern which
prevents estimating B modes even in absence of noise. To check whether this
also applies to our component 
separation pipeline, we estimated the CMB B-mode power spectrum from the CMB
maps recovered by the CCA in the noiseless case. It is important to note,
however, that this cannot be really considered as the application of our
pipeline to an ideal (noiseless) experiment because, while the component
reconstruction has been made by setting to zero the noise term in
eq.~(\ref{eq:quarantadue}), 
the spectral parameters going into the reconstruction matrix used to perform
component separation are those  estimated by assuming the real pre-flight
estimates of {\sc Planck} noise level for the nominal mission duration (14
months). The errors on foreground spectral parameters, and therefore those on
the recovered CMB map, are thus substantially larger than those expected for
an ideal noiseless experiment. 

\begin{figure}
\begin{center}
\includegraphics[width=8cm,keepaspectratio]{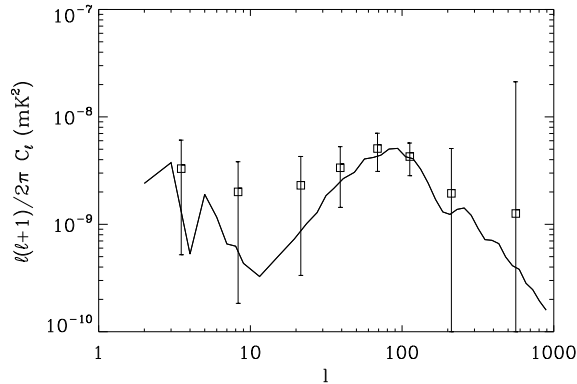}
\caption{Recovered B-mode power spectrum in the noiseless case (squares with 1$\sigma$ component separation error bars) compared with the input model (solid line).\label{fig:bb_sigonly}}
\end{center}
\end{figure}

The results are  shown in Fig.~\ref{fig:bb_sigonly}. For the tensor to scalar
ratio adopted in the simulations ($r=0.1$), our component separation method
proves to be capable of recovering the B-mode power spectrum over a quite
broad multipole range without any noticeable spurious feature due to residual
foreground contamination or to cross-correlation with the E-mode. Note that
the error bars due to  component separation shown in
Fig.~\ref{fig:bb_sigonly} are estimated \emph{in the presence of noise} while
the CMB recovery is made using noiseless maps. Since the noise dominance will
be much stronger for the B-mode than for the E-mode, the subtraction of the
noise bias will be a very delicate and computationally demanding, but
conceptually easy, operation. 

%% file: sec9.tex
We presented and tested on realistically simulated {\sc Planck} polarization data a pipeline for component separation based on the CCA method \citep{bedini2005,bonaldi2006,wmapbonaldi}.
This method exploits the data statistics to estimate the frequency behaviour of the diffuse components superimposed to the CMB, described in terms of a limited set of parameters, which in our case are synchrotron and thermal dust spectral indices.

The most recent harmonic domain version of CCA proved to be even superior to
the original pixel domain one, which however gave very good results for
simulated {\sc Planck}  \citep{bonaldi2006, leach08} and WMAP data \citep{wmapbonaldi} temperature data. As another improvement, we worked out a method to combine the results obtained with CCA on several regions of the sky to provide spatially-varying maps of spectral indices, with rms errors of only 0.015 for the dust component and of 0.08 for the synchrotron component, exploiting polarization data only ({\sc Planck} plus the WMAP 23 GHz channels).

Perhaps the most important new result of this paper is the elaboration, for the first time, of successful methods for estimating component separation errors. We presented two alternative error estimation methods for the CCA, relying on completely different assumptions (one is based on studying the marginal probability for each parameter estimated, the other on the redundancy of results for different patches of the sky) so that they can also be used to cross-check the results. A first application of these methods allowed us to identify the most uncertain values of the parameters and to compute the appropriate weights to be used to combine the estimates in different regions of the sky to build a smooth map of foreground spectral indices.


The components were reconstructed with a Generalized Least Square (GLS) solution in pixel space, which allowed us to fully exploit the spatially-varying information obtained.
Even  if the choice of the channels used in the reconstruction is optimized for the CMB, we were able to reconstruct a very accurate dust polarization map (correlation coefficient $\sim 1$ with the input maps). Errors in the estimation of the foreground spectral indices were successfully propagated to the foreground maps. They turned out to be well below those due to instrumental noise, except for the dust component close to the Galactic plane.

As for the CMB, we used ROMAster to estimate the polarization E-mode power spectrum from the reconstructed CMB map and found negligible effects by the residual foreground contamination even masking only 10\% of the sky. Again the component separation errors, propagated to the power spectrum, were found to be subdominant with respect to noise. We also showed that our component separation method can be useful to tackle B-mode detection, once the experimental noise level allows it.

\section*{Acknowledgements}
This research used resources of the National Energy Research Scientific Computing Center, which is supported by the Office of Science of the U.S. Department of Energy under Contract No. DE-AC02-05CH11231. Work supported by ASI through ASI/INAF Agreement I/072/09/0 for
the Planck LFI Activity of Phase E2.
SR and AB thank the PSM community of developers for useful discussions.



